%% file: main.tex
\documentclass[%
 aip,
 amsmath,amssymb,
 reprint, floatfix
]{revtex4-2}

\usepackage{graphics}
\usepackage{graphicx}
\usepackage{dcolumn}
\usepackage{bm}
\usepackage{MnSymbol}
\usepackage{siunitx}

\newcommand{\e}{\mathrm{e}}
\newcommand{\diff}{\mathrm{d}}
\newcommand{\kB}{k_\mathrm{B}}

\newcommand{\mbX}{\mathbf{X}}

\newcommand{\ba}{\beta_\mathrm{a}}
\newcommand{\RTg}{RT \gamma_\mathrm{a}}

\begin{document}

\title[Density and acoustic virials of helium]{Third density and
  acoustic virial coefficients of helium isotopologues from {\em ab initio}
  calculations} 

\author{Daniele Binosi}
\email{binosi@ectstar.eu}
 \affiliation{
   European Centre for Theoretical Studies in Nuclear Physics and
        Related Areas (FBK-ECT*), Trento, I-38123, Italy.}

\author{Giovanni Garberoglio}
\email{garberoglio@ectstar.eu}
 \affiliation{
   European Centre for Theoretical Studies in Nuclear Physics and
        Related Areas (FBK-ECT*), Trento, I-38123, Italy.}

\author{Allan H. Harvey}
\email{allan.harvey@nist.gov}
\affiliation{Applied Chemicals and Materials Division, National Institute of
  Standards and Technology, Boulder, CO 80305, USA.}

\date{7 May 2024}

\begin{abstract}
Improved two-body and three-body potentials for helium have been used to calculate from first
principles the third density and acoustic virial coefficients for both $^4$He and $^3$He.  For the
third density virial coefficient $C(T)$, uncertainties have been reduced by a factor of 4--5
compared to the previous state of the art; the accuracy of first-principles $C(T)$ now exceeds that
of the best experiments by more than two orders of magnitude.  The range of calculations has been
extended to temperatures as low as 0.5~K.  For the third acoustic virial coefficient
$\gamma_\mathrm{a}(T)$, we applied the Schlessinger Point Method, which can calculate
$\gamma_\mathrm{a}$ and its uncertainty based on the $C(T)$ data, overcoming some limitations of
direct path-integral calculation.  The resulting $\gamma_\mathrm{a}$ are calculated at temperatures
down to 0.5~K; they are consistent with available experimental data but have much smaller
uncertainties.  The first-principles data presented here will enable improvement of primary
temperature and pressure metrology based on gas properties.
\end{abstract}

\maketitle

\tableofcontents

\section{Introduction}

In recent years, great advances in gas-based pressure and temperature
metrology have been enabled by the ability to calculate properties of small
numbers of helium atoms from first principles with much smaller
uncertainties than they can be measured.  When combined with the ability to
make highly precise electromagnetic or acoustic measurements on gas
samples, the absolute temperature or the thermodynamic pressure can now be
measured with smaller uncertainties than previously possible.  The primary
methods for state-of-the-art temperature metrology are acoustic gas
thermometry,\cite{Moldover_2014} dielectric-constant gas
thermometry,\cite{Gaiser_2015} and refractive-index gas
thermometry.\cite{Rourke_2019, Rourke_2021} We also note the recent
development of a primary pressure standard based on dielectric measurements
of helium.\cite{Gaiser20,Gaiser_2022} A recent review describes the
contributions of first-principles calculated gas properties for precision
temperature and pressure metrology.\cite{Review_2023}

A key component of these efforts is the calculation of the virial
coefficients that describe the deviation from ideal-gas behavior.  For a
gas of molar density $\rho$ at temperature $T$, the virial expansion is
\begin{equation}
  \frac{p}{\rho R T} = 1 + B(T) \rho + C(T) \rho^2 + \cdots,
  \label{eq:virial}
\end{equation}
where $p$ is the pressure and $R$ is the molar gas constant.  The second
virial coefficient $B(T)$ depends on the interaction between two molecules,
the third virial coefficient $C(T)$ depends on interactions among three
molecules, and so on.

The state-of-the-art pair potential for helium incorporates higher-order
effects (adiabatic correction to the Born--Oppenheimer approximation,
relativistic effects, quantum electrodynamics) to produce extremely small
uncertainties in the potential over the entire physically relevant range of
distances.\cite{u2_2020} It has been used to calculate values of $B(T)$
over a wide temperature range; this calculation benefits from the fact that
an exact quantum calculation of $B(T)$ can be performed with a phase-shift
method.  The relative uncertainty of $B(T)$ for $^4$He near room
temperature is now on the order of $10^{-5}$, and the uncertainties for
$^3$He are similar.\cite{u2_2020}

For the third virial coefficient $C(T)$, no exact quantum solution is
known, but the path-integral Monte Carlo (PIMC) method can be used to
incorporate quantum effects with an accuracy limited only by the available
computing resources.  The last comprehensive first-principles calculations
of $C(T)$ were published by Garberoglio \textit{et al.} in
2011.\cite{Garberoglio2011a,Garberoglio2011}
These calculations used the pair potential of Cencek \textit{et
  al.},\cite{u2_2010} which has been superseded by the 2020 work of
Czachorowski \textit{et al.}\cite{u2_2020} that yields uncertainties in $B$
smaller by a factor of 5--10.  It also used the three-body potential
reported by Cencek \textit{et al.},\cite{FCI} which has recently been
improved upon by the work of Lang \textit{et al.}~\cite{u3_2023}

At low temperatures, the uncertainty of the 2011 values of $C(T)$ was
dominated by the convergence of the PIMC calculations; limitations on
computing resources resulted in a lower limit of 2.6~K for the calculated
results.  At higher temperatures, the largest contribution to the
uncertainty of $C(T)$ was that due to the uncertainty of the three-body
potential.

We are now in a position to improve on the 2011 calculations in several
ways.  The uncertainty due to the potentials will be greatly reduced by
using state-of-the-art two-body\cite{u2_2020} and three-body~\cite{u3_2023}
potentials; the new potentials reduce that component of our uncertainty by
approximately a factor of 5 throughout the entire temperature range.  We are
able to reduce the statistical uncertainty from the PIMC calculation not
only through increased computing power, but also by an improved propagator
that accelerates the PIMC convergence.  This allows us to compute values of
$C(T)$ down to 0.5~K, with the statistical uncertainty only becoming the
dominant uncertainty contribution below 2~K.  We have also developed a more
rigorous method for estimating the component of the uncertainty in $C(T)$
that results from uncertainties in the potentials used.

A related quantity of interest is the third acoustic virial coefficient
$\gamma_\mathrm{a}$, which arises in a low-density expansion [similar to
  Eq.~(\ref{eq:virial})] for the sound speed around its ideal-gas value and
is essential for acoustic gas thermometry.  In the 2011
work,\cite{Garberoglio2011} slow convergence of the PIMC calculations
limited the accuracy attainable for $\gamma_\mathrm{a}$.  In this work, we
introduce novel computational methods that enable a significant reduction
of the uncertainty of the calculation of the acoustic
virials. Nevertheless, our results are still limited by the statistical
uncertainty of the Monte Carlo calculations for $T \leq 100$~K.  We also present a novel
method of deriving $\gamma_\mathrm{a}(T)$ from $C(T)$ data, and we argue that it provides
an upper-bound estimate of the propagated uncertainty.

\section{Ab-initio calculation of virial coefficients}

The calculation of virial coefficients has been performed using the latest
pair~\cite{u2_2020} and three-body~\cite{u3_2023} potentials for
helium. These potentials come with well-defined uncertainty estimates,
which we will assume to correspond to an expanded $k=2$ uncertainty. We
note that, due to the inclusion of relativistic effects, the pair potential
for ${}^3$He is slightly different from that for ${}^4$He.  In principle, a
similar difference exists for the three-body potential, but this difference
would be much smaller than the uncertainty in the potential so we use the
three-body potential derived for ${}^4$He in Ref.~\onlinecite{u3_2023} for
both isotopes.

In previous calculations, it was noted that the largest contribution to the
uncertainty of the third virial coefficient was due to the propagated
uncertainty of the three-body potential.~\cite{Garberoglio2009b,
  Garberoglio2011} The higher accuracy of the three-body
potential used in this work, which resulted in uncertainties reduced by 3
to 5 times with respect to the previous potential,~\cite{u3_2023} required
us to develop some improved approaches for the calculation of the virial
coefficients. Since the general framework is still the same as in our previous
work,~\cite{Garberoglio2011a,Garberoglio2011,Review_2023} we will briefly describe
the new methods developed for the present work.

\subsection{Uncertainty budget}

Before presenting the main results of this paper, it is worth examining the
various contribution to the uncertainty budget of our calculations.
In general, there are three sources: the statistical uncertainty of the
path-integral Monte Carlo calculations, and the propagated uncertainties
from the pair and three-body potentials. The latter two contributions can be
further separated into contributions from Boltzmann statistics
(indistinguishable particles, which is the leading contribution to $C(T)$
at $T > 4$~K), and from the odd and even exchanges, which depend on the
bosonic or fermionic nature of the isotope under consideration and become
relevant at low temperatures.

In our earlier works, we used two methods to propagate the uncertainty from
the potentials to the third virial coefficient. In the
first,~\cite{Garberoglio2009b,Garberoglio2011,Garberoglio2011a} we would
compute the virial coefficients with modified potentials (adding and
subtracting their uncertainty) and compute the uncertainty from the difference
between these values.
This approach is reasonably good at high temperatures, but fails at the
lowest ones because the statistical uncertainty of the Monte Carlo
calculation becomes large, and one needs to perform very computationally
intensive calculations to reduce it significantly. Additionally, a rigid shift of the pair potential
in the case of virial coefficients higher than the second may result in a positive, negative, or zero
shift in the virial coefficient, thus calling into question on the validity of this approach.

Recently, we developed an alternative method that evaluates the
propagated uncertainty by performing functional differentiation of the
formula for $C(T)$ with respect to the potentials.~\cite{Garberoglio2021a}
We used a semiclassical approach using the fourth-order Feynman--Hibbs effective potential to
evaluate the propagated uncertainty, which we deemed adequate for the scope of our previous
calculations.

In this work, we extend this latter method by taking the functional
derivative directly on the path-integral expression for
$C(T)$. Consequently, our expression for the propagated uncertainty from
the potential is valid at all temperatures. It turns out that the
computational effort to determine the propagated uncertainties using this
approach is much smaller than that needed to compute virial
coefficients.
Hence, we could evaluate the uncertainty in $C(T)$ propagated from the potentials. Subsequently, we
conducted extensive Monte Carlo simulations aimed at reducing the statistical error to a level below the
uncertainty attributed to the potentials.

\begin{figure}
\center\includegraphics[width=0.9\linewidth]{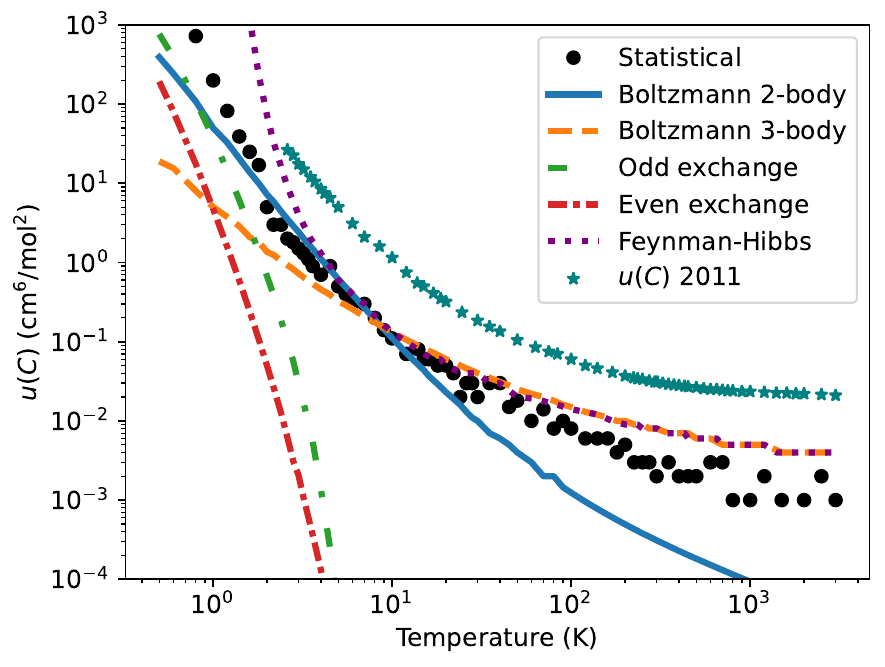}  
\caption{Uncertainty budget for $C(T)$ of ${}^4$He in
the present work.}
\label{fig:uC_He4}  
\end{figure}

An overview of the uncertainty budget in the case of ${}^4$He is shown in
Fig.~\ref{fig:uC_He4}. The situation for ${}^3$He is very similar
and is not reported here.

First of all, we can retrospectively gauge the validity of the Feynman--Hibbs fourth-order
semiclassical estimation of the propagated uncertainty using the functional differentiation
approach. Inspection of Fig.~\ref{fig:uC_He4} shows that this approximation is quite good for $T
\geq 4$~K. At lower temperatures, the uncertainty obtained with the semiclassical approach generally
exceeds that obtained using the rigorous path-integral estimation and tends to increase quite
rapidly.

We also compare the uncertainty with the new potentials to the uncertainty of our previous
calculation, reported with starred symbols.~\cite{Garberoglio2011} As already pointed out in
Ref.~\onlinecite{u3_2023}, the accuracy of the new potentials, especially the three-body surface,
results in a reduction of the uncertainty by roughly a factor of 5 (and sometimes more) at all
temperatures above 2.6~K.

Additionally, the increased accuracy of the potential energy surfaces
required us to develop improved calculation methods, in order to be able
to reduce the statistical Monte Carlo uncertainty below that propagated
from the potentials. As will be described below, we used enhanced
propagators in the calculation of $C(T)$ and a novel approach based on the
idea of the {\em virial} estimator of the kinetic energy in path-integral
Monte Carlo calculations to reduce the uncertainty of $\RTg$.

In the case of $C(T)$, this approach was successful for $T \geq 2$~K, where
the uncertainty budget is dominated by the uncertainty propagated from the
potentials. At lower temperatures, the uncertainty budget is dominated by
the Monte Carlo contribution.

\begin{figure}
\center\includegraphics[width=0.9\linewidth]{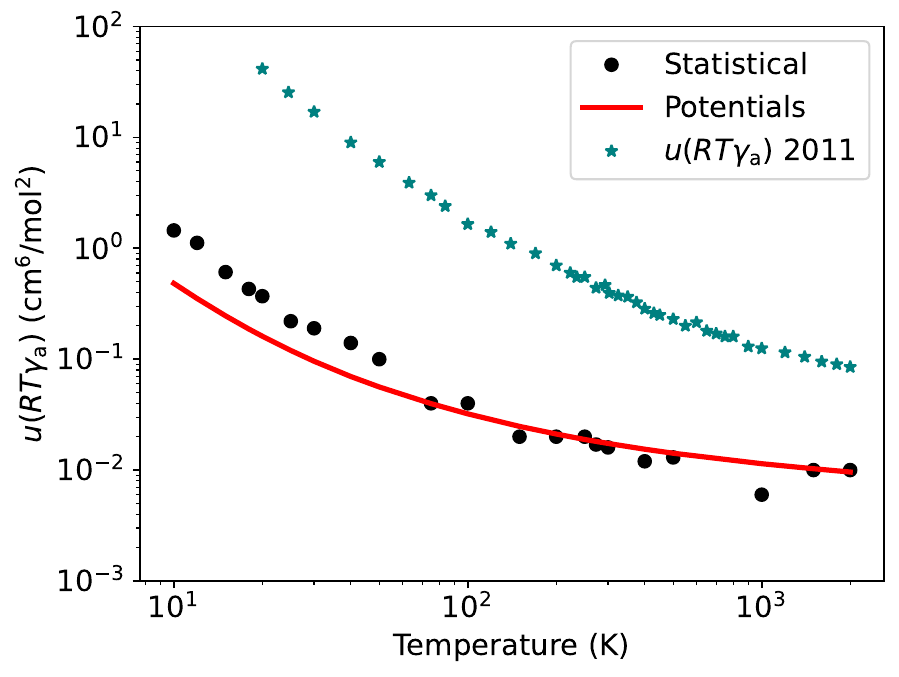}
  \caption{Uncertainty budget for $\RTg (T)$ of ${}^4$He in
  the present work.}
\label{fig:uRTg_He4}  
\end{figure}

The uncertainty budget for the third acoustic virial coefficient, $\RTg$, is reported in
Fig.~\ref{fig:uRTg_He4}. Also in this case, the propagation of the uncertainty from the pair and
three-body potentials has been performed by functional differentiation of the path-integral
expression for $\RTg$, which is reported in the Supplementary Material.  Compared with our previous
results,~\cite{Garberoglio2011} we can see that the combination of more accurate potentials and
reduced-variance estimators resulted in a reduction of the uncertainty by more than one order of
magnitude. However, despite our computational efforts, the statistical uncertainty still dominates
the budget at temperatures $T \leq 100$~K, up to a factor between 2 and 3.  In order to overcome this
limitation in the direct calculation of $\RTg$, we used the statistical Schlessinger Point Method
(SPM) to derive the third acoustic virial and its uncertainty directly from the results for
$C(T)$. This approach, as we will detail below, provides us with the most accurate estimation of
$\RTg$ and its uncertainty at low temperatures.

\subsection{The third density virial coefficient}

The virial expansion of Eq.~(\ref{eq:virial}) is a rigorous result of quantum statistical mechanics, which
also provides an exact formula relating $C(T)$ and the interaction among three particles.  $C(T)$ is
conveniently evaluated using the path-integral formulation of quantum statistical mechanics that
enables rewriting the expression for $C(T)$ involving three quantum particles into an equivalent
classical expression involving three ring polymers of $P$ beads each; the correspondence is exact in
the $P \to \infty$ limit.~\cite{FH65,Garberoglio2011a}

This equivalence is based on the Trotter--Suzuki identity, which -- in the
case of a quantum Hamiltonian $H = T+V$, where $T$ is the kinetic energy
and $V$ is the potential energy -- is written as
\begin{equation}
  \e^{-\beta (T + V)} =
  \lim_{P \to \infty}
  \left( \e^{-\beta T/P} \e^{-\beta V/P}\right)^P.
  \label{eq:TS}
\end{equation}
In actual calculations, one uses a finite value of $P$, which has to be
taken large enough so that the results obtained are converged to within a
specified uncertainty. For the calculation of virial
coefficients, it is found that the optimal value of $P$ is inversely
proportional to the temperature and that $P T \sim 2400$~K for helium.
The straightforward use of Eq.~(\ref{eq:TS}) implies that calculations at
lower temperature become progressively more demanding.

One way to overcome this difficulty is to develop more accurate
approximations of the right-hand side of Eq.~(\ref{eq:TS}). This idea was
first put forward by Takahashi and Imada~\cite{TI84} and subsequently
developed by Kono {\em et al.}~\cite{KTL88} and by Li and
Broughton.~\cite{LB87}
The latter authors showed that a more accurate and effective approximation
of the Trotter--Suzuki expansion is
\begin{eqnarray}
  \e^{-\beta(T+V)/P} &\sim& \e^{-\beta T / P} \e^{-\beta V /P}
  \e^{-(\beta/P)^3 W/24}
  \label{eq:LB}
  \\
  W &=& [[V,T],V] = \frac{\hbar^2}{m} \left| \nabla{V} \right|^2.
  \label{eq:W}
\end{eqnarray}
In the case of three particles, the potential $V$ can be expressed as a
function of the distances between the pairs, that is $r_{12}$, $r_{13}$, and
$r_{23}$, as
\begin{equation}
  V = U_2(r_{12}) + U_2(r_{13}) + U_2(r_{23}) + U_3(r_{12},r_{13},r_{23}),
\end{equation}
where $U_2(r)$ is the pair potential and $U_3(r_{12},r_{13},r_{23})$ is
the non-additive part of the three-body potential.
In this case, the squared gradient appearing in Eq.~(\ref{eq:W})
can be written as
\begin{widetext}
\begin{equation}
  \left|\nabla V\right|^2 =
  2 \left| \frac{\partial V}{\partial r_{12}} \right|^2 +
  2 \left| \frac{\partial V}{\partial r_{13}} \right|^2 +
  2 \left| \frac{\partial V}{\partial r_{23}} \right|^2 + 
  \frac{\partial V}{\partial r_{12}} \frac{\partial V}{\partial r_{13}} \cos\theta_1 +
  \frac{\partial V}{\partial r_{12}} \frac{\partial V}{\partial r_{23}} \cos\theta_2 +
  \frac{\partial V}{\partial r_{13}} \frac{\partial V}{\partial r_{23}} \cos\theta_3,
\end{equation}  
\end{widetext}
where $\theta_i$ is the internal angle at particle $i$ in the triangle made
by the three particles.

In the actual calculations, it has been found convenient to compute $C(T)$
as the sum of two parts: the first is obtained considering pair potentials
only (that is, assuming $U_3 = 0$), whereas the second part is the
contribution due to the non-zero value of
$U_3$.~\cite{Shaul2012,Garberoglio2021a} In general, the most time-consuming part of the calculation
involves the contribution from the pair 
potential, whereas the non-additive contribution to $C(T)$ is much less
computationally demanding. Therefore, we have used the standard {\em
  primitive} approximation~\cite{Ceperley1995} of Eq.~(\ref{eq:TS}) for the
latter contribution, and the Li--Broughton approximation of
Eq.~(\ref{eq:LB}) for the pair contribution only.  In this latter case, we found
that we could reach well-converged results using $P = \mathrm{nint}(4 +
\sqrt{120~\mathrm{K}/T})$ and $P = \mathrm{nint}(4 +
\sqrt{160~\mathrm{K}/T})$ for ${}^4$He and ${}^3$He, respectively.
The function $\mathrm{nint}(x)$ denotes the nearest integer to $x$.
The values of $P$ needed using the Li--Broughton approach should be compared with the values $P = \mathrm{nint}(4 +
2400~\mathrm{K}/T)$ that are needed, irrespective of the isotope, to reach convergence using the
primitive approximation, and that was used to compute the non-additive contribution to
$C(T)$. In the case of the Li--Broughton approximation, a much smaller number of beads is
needed to reach convergence in the path-integral results. This more than offsets the additional
calculations needed to evaluate the quantity $W$ of Eq.~(\ref{eq:W}).

At low temperatures, quantum statistical effects contribute to the value of
the third virial coefficient. We evaluated these contributions using the
primitive approximation, which is the same approach adopted in
Ref.~\onlinecite{Garberoglio2011a}.

\subsection{The third acoustic virial coefficient}

The acoustic virial coefficients appear in the pressure expansion of the
speed of sound $u$, according to
\begin{equation}
  u^2 = \frac{\gamma_0 RT}{M} \left[ 1 + \ba \frac{p}{RT} +
  \gamma_\mathrm{a} \frac{p^2}{RT} + \ldots \right],
\end{equation}
where $\gamma_0 = 5/3$ for a monoatomic gas and $M$ is the molar mass.  $\ba$ and
$\RTg$ can be calculated from the first and second temperature derivatives of the second and third
density virial coefficients according to the formulae\cite{Gillis96}
\begin{eqnarray}
  \ba(T) &=& \underbrace{2B}_{1} +
  \underbrace{2(\gamma_0-1) T \frac{\diff B}{\diff T}}_{2} +
  \underbrace{\frac{(\gamma_0 -1)^2}{\gamma_0} T^2 \frac{\diff^2B}{\diff T^2}}_{3},
  \label{eq:ba} \\
  Q &=& B +
  (2\gamma_0-1) T \frac{\diff B}{\diff T} +
  (\gamma_0-1) T^2 \frac{\diff^2B}{\diff T^2}
  \label{eq:Q} \\
  \RTg &=& \frac{\gamma_0 -1}{\gamma_0} Q^2 - \ba(T) B(T) + 
  \frac{2 \gamma_0 + 1}{\gamma_0} C + \nonumber \\
  & &
  \frac{\gamma_0^2-1}{\gamma_0} T \frac{\diff C}{\diff T} +
  \frac{(\gamma_0 -1)^2}{2 \gamma_0} T^2 \frac{\diff^2C}{\diff T^2},
  \label{eq:RTg}
\end{eqnarray}
where we have indicated three terms ($1$, $2$, and $3$) in $\ba$ for later convenience.

As noted in Ref.~\onlinecite{Garberoglio2011} the
path-integral expression used for the calculation of $\RTg$ involves, due
to the presence of temperature derivatives, expressions analogous in form
to the so-called {\em thermodynamic} kinetic energy estimator, which is
known to have a large variance.~\cite{Tuckerman10} As a consequence, in the
calculations of $\RTg$ the largest part of the uncertainty was due to the
statistical uncertainty of the Monte Carlo calculations.

In order to reduce this effect as much as possible, we developed a new
approach based on the same ideas that led to the {\em virial} estimator of
the kinetic energy.~\cite{Herman82}
The lengthy derivations of the path-integral formulae leading to a reduced-variance
estimation of the third acoustic virial coefficient are reported in the
Supplementary Material.
However, as we discuss in the following, even these improved formulae,
which resulted in a reduction of the statistical uncertainty in the
path-integral Monte Carlo evaluation of $\RTg$ by more than one order of magnitude
at $T=20$~K, were not enough to reduce the statistical uncertainty below that propagated from the
potentials at the lowest temperatures investigated in this work.

In order to provide more accurate estimates of the acoustic virial coefficients, we employ the
Schlessinger Point Method (SPM). The SPM allows us to compute the first and second temperature
derivatives, and their uncertainties, directly from our calculated $C(T)$ data, removing the need to
simulate the acoustic virials.

\section{\label{sec:SPM}Statistical Schlessinger Point Method}

\subsection{Description of the method}

Let us denote by $\mathsf{D}_N$ the set of all the $N$ computed pairs of some virial coefficient
$v_i$ associated with a given temperature $T_i$: 
\begin{align}
	\mathsf{D}_N = \{(T_i , v_i = v(T_i)),\ i=1,\dots, N\}.
	\label{DNexact}
\end{align} 
Within a subset $\mathsf{D}_M\subseteq\mathsf{D}_N$ (with $M<N$), one can construct the Schlessinger Point Method (SPM) continued fraction interpolator~\cite{Schlessinger:1968}
\begin{align}
	I_M(T)=\cfrac{v_1}{1+\cfrac{a_1(T-T_1)}{1+\cfrac{a_2(T-T_2)}{1+\cfrac{\cdots}{1+\cfrac{\cdots}{a_{M-1}(T-T_{M-1})}}}}},	
		\label{SPMcf}
\end{align}
where the $M-1$ coefficients $a_i$ are recursively determined from the formulas
\begin{subequations}
\begin{align}
	(T_2-T_1)a_1&=v_1/v_2-1,\\
	(T_\ell-T_{\ell+1})a_\ell&=1+\cfrac{a_{\ell-1}(T_{\ell+1}-T_{\ell-1})}{1+\cfrac{a_{\ell-2}(T_{\ell+1}-T_{\ell-2})}{1+\cfrac{a_{\ell-3}(T_{\ell+1}-T_{\ell-3})}{1+\cfrac{\cdots}{1+\cfrac{a_1(T_{\ell+1}-T_1)}{1-v_1/v_{\ell+1}}}}}},
\end{align}
\end{subequations}
and are such that \mbox{$I_M (T_i ) = T_i$}, $ \forall\ T_i\in \mathsf{D}_M$. 

The interpolator~\eqref{SPMcf} can be cast in the rational form 
\begin{align}
	I_M(T)=\frac{P_M(T)}{Q_M(T)},
	\label{rexp}
\end{align}
where $P_M$ and $Q_M$ are polynomials whose degree is determined by the size $M$ of the subset
$\mathsf{D}_M$ chosen: $(M-1)/2$ (both $P_M$ and $Q_M$) if $M$ is odd; $M/2-1$ ($P_M$) and $M/2$
($Q_M$) if $M$ is even. Thus, for large $T$,  $I_M\sim \mathrm{const.}$ (respectively, $I_M\sim
1/T$) for $M$ odd (respectively, even).

While the SPM shares the same rational expression of Eq.~(\ref{rexp}) with a Pad\'e approximant, the
idea behind it is completely different. The latter is in fact defined as an expansion of a function
near a specific point: its coefficients are thus constructed from values of higher derivatives at
that point, so that the approximant's power series agrees with that of the original function. In
the case of SPM, Eq.~(\ref{rexp}) is rather constructed using the original function values at
\emph{different} points.\footnote{Sometimes this is also referred to as a multipoint-Pad\'e
expansion.} Thus, $I_M$ is able to capture the behavior of $v$ over a range of values extending
beyond that of $\mathsf{D}_M$ (and even $\mathsf{D}_N$) without the need to compute any
derivatives.

How well the interpolator of Eq.~(\ref{SPMcf}) is capable of reproducing the entire dataset
$\mathsf{D}_N$ -- and, correspondingly, the function $v$ -- mainly depends on the precision of the
starting dataset $\mathsf{D}_N$. For example, when exact numerical data are considered as
in Eq.~(\ref{DNexact}), $I_M$ is practically indistinguishable from $v$ independent of the number
$M$ of points in $\mathsf{D}_M$ (provided that the latter is large enough to capture the basic
features of the  curve $v$ we wish to describe). However, in the presence of uncertainties one has 
\begin{align}
	\mathsf{D}_N = \{(T_i , v_i, \epsilon_i),\ i=1,\dots, N\}.
	\label{DNerr}
\end{align} 
so that the $v_i$ are statistically distributed around the true curve $v$ with variance $\epsilon_i$. 

A veracious reconstruction of the function $v$ can be obtained also in this case if the SPM is
combined with resampling to propagate uncertainties. To this end, we generate from Eq.~(\ref{DNerr})
a replica set $\mathsf{D}^\mathrm{r}_N$ by randomly drawing from each $v_i$ in $\mathsf{D}_N$ a new one
$v^\mathrm{r}_i$ from a normal distribution ${\cal N}$ with a mean value equal to $v_i$ and a standard
deviation equal to its associated standard uncertainty $\epsilon_i$: 
\begin{align}
	\mathsf{D}^\mathrm{r}_N = \{(T_i , v^\mathrm{r}_i={\cal N}(v_i, \epsilon_i)),\ i=1,\dots, N\}.
	\label{DNrepl}
\end{align} 
Next, fixing $M$ at a suitable value,
\footnote{The SPM results have been proven to be independent of the number of input points, see {\it
  e.g.}, Ref.~\onlinecite{Cui:2021vgm}. Herein we always set $M=20$ for the interpolation of both
the second virial coefficient $B$ (for which a total of $N=122$ values were available) and the third
one $C$ (for which a smaller dataset of $N=68$ values was available).}  we randomly choose the
subset $\mathsf{D}^\mathrm{r}_M\subseteq\mathsf{D}^\mathrm{r}_N$ and proceed to construct the
corresponding SPM interpolator $I_M^\mathrm{r}$ from Eq.~(\ref{SPMcf}). Repeating these steps for a
sufficiently large number of replicas $n^\mathrm{r}$ gives rise to a large population of
interpolators that can be filtered according to suitable criteria.

More specifically, we require that the interpolator $I_M^\mathrm{r}$ is such that: {\it i}) it is
smooth (continuous with all its derivatives, $I_M^\mathrm{r}\in C^\infty$) on the positive real
axis\footnote{This is equivalent to requiring that statistical fluctuations in the replica subset do
not cause the denominator $Q_M^\mathrm{r}$ appearing in Eq.~(\ref{rexp}) to develop real zeros on
$\mathbb{R}_{>0}$.}  ${\mathbb R}_{>0}$; {\it ii}) it deviates less than $\epsilon_i$ from 95\% of
the remaining $N-M$ points in $D_N$.

For the specific cases of the second and third virial coefficients studied here, these conditions
are very stringent: using the data computed in Ref.~\onlinecite{u2_2020} for $B$, only $\sim0.002\%$
(${}^3$He) and $\sim0.01\%$ (${}^4$He) of the constructed interpolators satisfied them; and those
percentages drastically drop for $C$ using the data computed in this paper (as in this case
$\mathsf{D}_N$ contains fewer data with larger uncertainties), where one has a mere~0.0003\%
(further down to 0.00001\%) in the ${}^3$He (${}^4$He) cases, respectively.

Each of the derived curves can be utilized to construct a smooth description of the second
($\ba$) and third ($\RTg$) acoustic virial coefficients by directly applying
Equations~(\ref{eq:ba}) and~(\ref{eq:RTg}). However, this process involves calculating up to
second-order derivatives of the constructed interpolators, which may introduce spurious oscillations
into the resulting curves.
The curves displaying these fluctuations are removed from the final set of interpolators.
Thus, one obtains a different number of interpolators for the $B$ and $C$ viral coefficients in the
${}^4$He and ${}^3$He cases. However, as both $B$ and $C$ enter in the calculation of
$RT\gamma_{\mathrm{a}}$ the minimum number of available interpolators between $B$ and $C$ is retained:
more specifically, one ends up with 587 SPM interpolators for the ${}^4$He $B$ and $C$ virial
coefficients and 378 for the ${}^3$He ones. 

This continuous framework allows for the evaluation of virial coefficients at any specified
temperature. Here, the SPM output represents the average of the interpolator curve values at the
given temperature, with the uncertainty derived straightforwardly from the standard deviation of
these values. 

\begin{figure}[t]
  \center\includegraphics[width=0.8\linewidth]{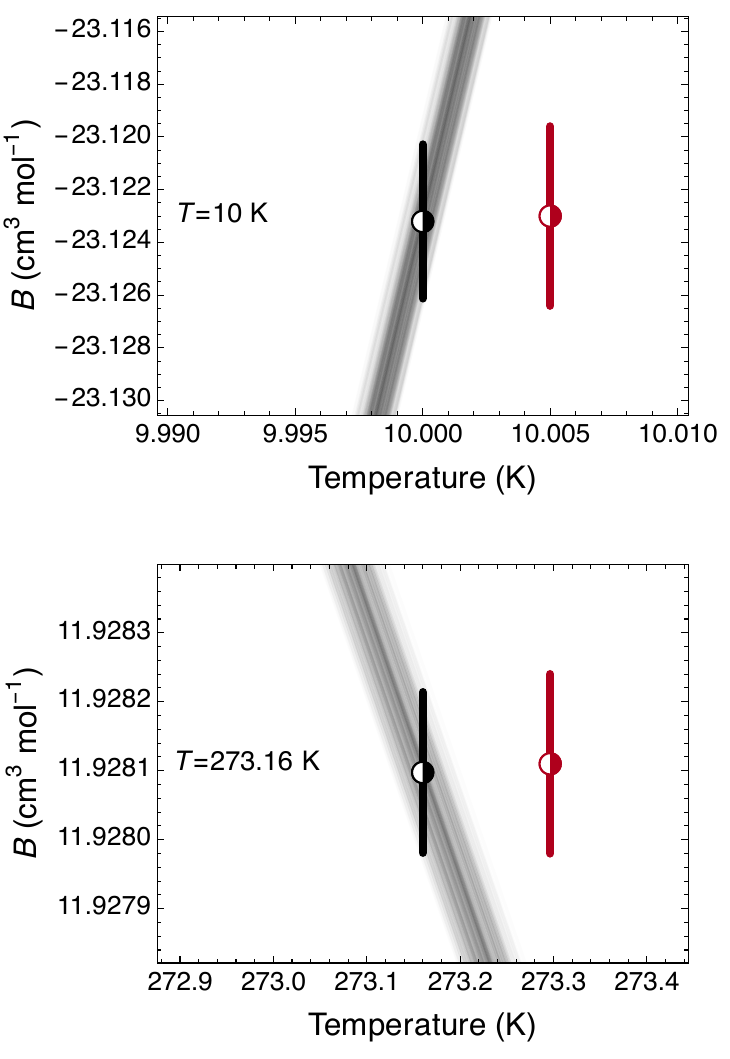}  
  \caption{Interpolating curves generated using the SPM method in the case of $B(T)$ for
    ${}^4$He at $T=10$~K (upper panel) and $T=273.16$~K (lower panel). The black symbol denotes the
    average value and $k=2$ expanded uncertainty derived from the SPM curves. The red symbol, which
    has been displaced to the right for the sake of clarity, reports the average value and expanded
    uncertainty from the original calculation.~\cite{u2_2020} }
\label{fig:B_SPM}  
\end{figure}

\subsection{The second acoustic virial coefficient}
\label{sec:beta_SPM}

To examine the capability of the SPM method in estimating uncertainty propagation for
acoustic virials, we investigated in detail its performance in the case of the second acoustic
virial coefficient $\ba(T)$ of ${}^4$He from $B(T)$ data, comparing it with the direct calculation
using the phase-shift method.~\cite{u2_2020}

We first present in Fig.~\ref{fig:B_SPM} the SPM interpolators that reproduce $B(T)$ and its
uncertainty at two temperatures (additional illustrative plots are available in the Supplementary
Material). One can see that the set of SPM curves reproduces quite well the average value and
expanded uncertainty of $B(T)$ across a wide temperature range.

\begin{figure}
  \center\includegraphics[width=0.8\linewidth]{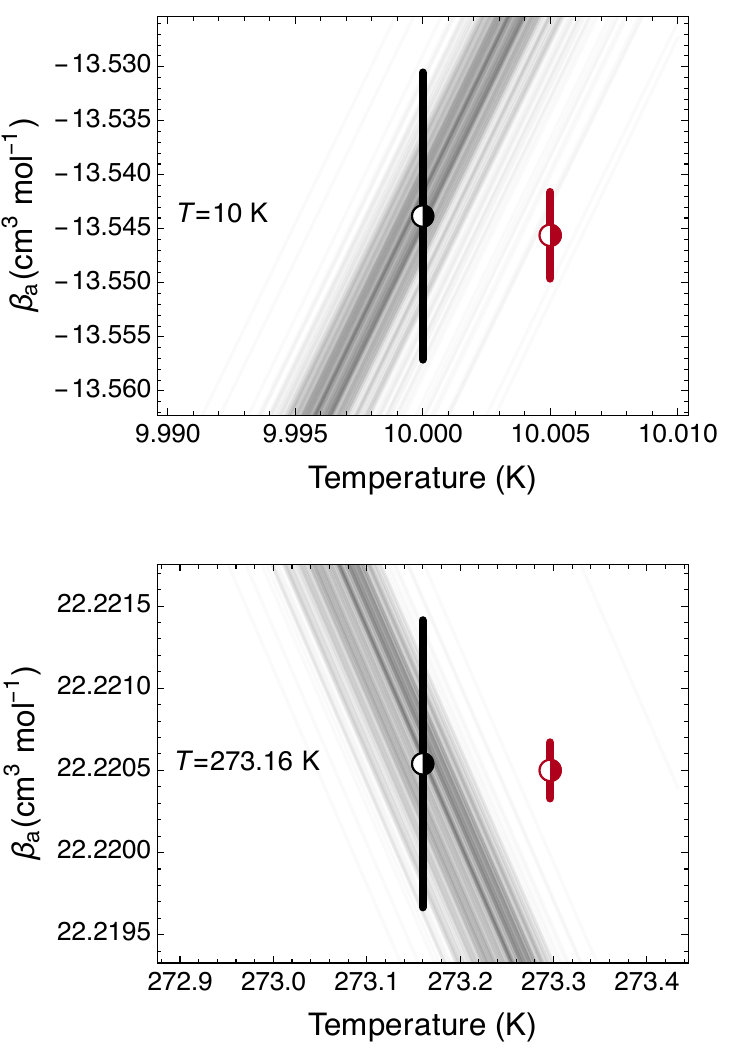}  
  \caption{Interpolating curves generated using the SPM method in the case of $\ba(T)$ for
    ${}^4$He at $T=10$~K (upper panel) and $T=273.16$~K (lower panel). The black symbol denotes the
    average value and $k=2$ expanded uncertainty derived from the SPM curves. The red symbol, which
    has been displaced to the right for the sake of clarity, reports the average value and expanded
    uncertainty from the original calculation.~\cite{u2_2020} }
\label{fig:beta_SPM}  
\end{figure}

The second acoustic virial coefficient $\ba(T)$ of ${}^4$He computed with SPM is shown in
Fig.~\ref{fig:beta_SPM}, where we also compare it with the direct calculations, at the same two
temperatures as in Fig.~\ref{fig:B_SPM} (also in this case, the plots for more temperatures are shown
in the Supplementary Material). We see that the SPM-derived values are in very good agreement
with the ones obtained by direct computation;~\cite{u2_2020} however, in this case the SPM approach
provides an uncertainty higher than that computed as the difference of the values of
$\ba(T)$ obtained with rigidly shifted potentials.

\begin{figure}
    \center\includegraphics[width=0.8\linewidth]{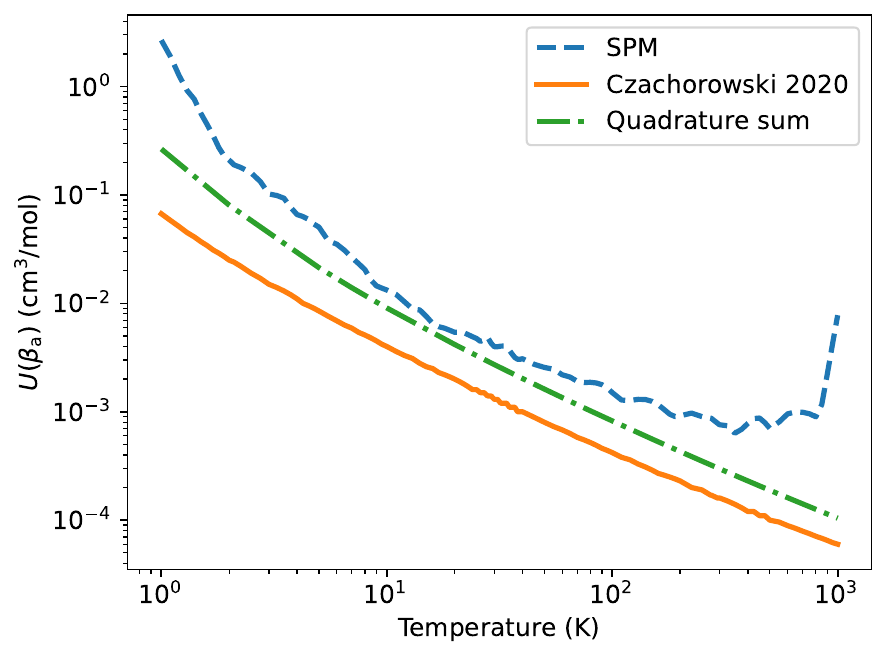}
  \caption{Expanded ($k=2$) uncertainty of $\ba(T)$ for
    ${}^4$He as a function of temperature. Solid line: estimate from
    Ref.~\onlinecite{u2_2020}, obtained by the difference between the values of $\ba$
    computed using rigidly shifted potentials. Dashed line: estimate from the SPM
    approach. Dot-dashed line: sum in quadrature of the uncertainty propagated from the three terms
    $1$, $2$, and $3$ in Eq.~(\ref{eq:ba}) using the functional differentiation approach.}
\label{fig:beta_SPM_T}  
\end{figure}

We show in Fig.~\ref{fig:beta_SPM_T} the ratio between the SPM estimated uncertainty of $\ba(T)$ and
the uncertainty computed in Ref.~\onlinecite{u2_2020}. We notice that SPM tends to significantly
overestimate the uncertainty at the boundaries of the temperature range considered. SPM estimates close to the boundaries could be improved by adding knowledge of the
limiting behavior of $B(T)$ ({\em e.g.}, a known power-law dependence on $T$), but we did not pursue
this further in this paper, because uncertainty propagation using PIMC methods can be performed
efficiently even with limited computational resources at high temperature, and there is presently no
theoretical treatment of exchange effects for acoustic virial coefficients, which are expected to
contribute significantly below $\approx 5$~K. Due to the aforementioned SPM behavior 
at the boundaries of the temperature range, we will report $\RTg$ and its uncertainty evaluated with
SPM down to $T= 1$~K (see Supplementary Material).

In the intermediate temperature region, however, the SPM uncertainty
estimates of $\ba$ are $3$ to $4$ times larger than those computed in Ref.~\onlinecite{u2_2020}.
The main reason for this result is that the SPM method is statistical in nature, and it is based on
generating a series of smooth curves that interpolate $B(T)$ values and their estimated
uncertainty. For each of these curves, one calculates the temperature derivatives and evaluates
$\ba(T)$ according to Eq.~(\ref{eq:ba}). The uncertainty of $\ba$ is evaluated from the standard
deviation of the distribution of the values of $\ba$ at each temperature.

However, inspection of Eq.~(\ref{eq:ba}) for the acoustic virial shows that it is obtained as the
sum of three terms ($1$, $2$, and $3$), each of which depends on the pair potential. When
considering the effect of the uncertainty of the pair potential on $\ba(T)$, these three terms will
most likely exhibit a correlated variation, which leads to cancellations since the contributions
have opposite signs due to the shape of the $B(T)$ function. This correlated behavior is not
taken into account by the SPM approach, which in fact provides a rigorous upper bound to the actual
propagated uncertainty.

Another way to arrive at a similar conclusion is to consider the three
terms delineated in Eq.~(\ref{eq:ba}). The approach used to propagate the
uncertainty in Ref.~\onlinecite{u2_2020} estimates it from
\begin{equation}
  U(\ba) = \frac{1}{2} \left|
  \ba(T; u_2 +\delta u_2) - \ba(T; u_2- \delta u_2)
  \right|,
  \label{eq:Uba}
\end{equation}
where $\ba(T; u)$ denotes the value of the second acoustic virial coefficient computed using the pair
potential $u$. This way of estimating uncertainty is based on a rigid shift of the pair
potential $u_2$ according to its estimated uncertainty $\delta u_2$ and it is not statistical in
nature. Notice also that it would include any correlation between the shifts of the three terms in
Eq.~(\ref{eq:ba}) induced by a variation of $\delta u_2$.

The effect of these correlations can be evaluated by computing the uncertainty of each of the three
terms in Eq.~(\ref{eq:ba}) using the functional derivative method and adding them in
quadrature. This procedure is tantamount to neglecting that all the terms would be indeed be
affected by the {\em same} variation $\delta u_2$ of the underlying pair potential, and in fact it
produces propagated uncertainties that are approximately {\em twice} as much as those estimated in
Ref.~\onlinecite{u2_2020}. The results of these calculations are reported as a dot-dashed line in
Fig.~\ref{fig:beta_SPM_T}.

From these considerations, we deduce that the SPM approach is a convenient way to propagate
uncertainties, and that it leads to an {\em upper bound} of the actual uncertainty that would be
obtained by either a rigid shift of the potential or, in the case of higher-order coefficients, by
the functional differentiation approach.~\cite{Garberoglio2021a,Review_2023}

\section{Results and discussion}

\subsection{Helium-4}

\subsubsection{Third virial coefficient, $C(T)$}

The values of $C(T)$ for ${}^4$He are reported in Tables~\ref{tab:CHe4_low} and
\ref{tab:CHe4_high}. At low temperatures, where quantum exchange effects are significant, $C(T)$ is
obtained as a sum of three terms:~\cite{Garberoglio2011a} the first corresponds
to Boltzmann statistics (distinguishable particles), and is the only term that contributes at high
temperatures, whereas the other two terms come from the odd and even exchange terms in the partition
function. Analogously to our previous findings, exchange effects are
appreciable only for temperatures $T \leq 6$~K.

Please note that we have slightly changed our notation for the exchange contribution
compared to Ref.~\onlinecite{Garberoglio2011a}. The definition of the Boltzmann component remained the same,
whereas contributions from the old quantity $C_\mathrm{B}$ have been included in the new definition
of $C_\mathrm{odd}$ and $C_\mathrm{even}$ which now include {\em all} the terms that involve an odd
or even permutation, respectively.\cite{Garberoglio2021a}
This change corresponds to how the various contributions are calculated: we found it convenient to
collect all similar terms in the same calculation in order to reduce the statistical
variance. Additionally, the odd and even contributions already contain the weights coming from
considering the nuclear spin $I$, so that we have $C(T) = C_\mathrm{Boltz} + C_\mathrm{odd} +
C_\mathrm{even}$ irrespective of the isotope.

\begin{turnpage}
\begin{table*}
  \caption{The third virial coefficient $C(T)$ and its expanded $(k=2)$
    uncertainty $U(C)$ for ${}^4$He at low temperatures. The other columns
    report the contributions due to Boltzmann statistics
    ($C_\mathrm{Boltz}$), and the odd ($C_\mathrm{odd}$) and even exchange
    ($C_\mathrm{even}$), together with their standard $(k=1)$ statistical
    uncertainties due to the Monte Carlo calculation. The last column
    reports the combined standard $(k=1)$ uncertainty propagated from the
    pair and three-body potentials. Temperatures are in units of K and $C(T)$ in units of cm${}^6$/mol${}^2$.}
  \begin{tabular}{d|d|d||d|d|d|d|d|d|d}
    \multicolumn{1}{c|}{$T$} &
    \multicolumn{1}{c|}{$C$} &
    \multicolumn{1}{c||}{$U(C)$} &
    \multicolumn{1}{c|}{$C_\mathrm{Boltz}$} &
    \multicolumn{1}{c|}{$u(C_\mathrm{Boltz})$} &
    \multicolumn{1}{c|}{$C_\mathrm{odd}$} &
    \multicolumn{1}{c|}{$u(C_\mathrm{odd})$} &
    \multicolumn{1}{c|}{$C_\mathrm{even}$} &
    \multicolumn{1}{c|}{$u(C_\mathrm{even})$} &
    \multicolumn{1}{c}{$u_\mathrm{pot}$} \\
    \hline
    \hline
0.5	&	-2112555	&	20631	&	-779262	&	5189	&	-599309	&	4837	&	-733983	&	7437	&	882	\\
0.6	&	-962869	&	6194	&	-395907	&	1497	&	-246498	&	1816	&	-320464	&	1959	&	466	\\
0.8	&	-273391	&	1476	&	-136279	&	376	&	-50458	&	395	&	-86654	&	473	&	155	\\
1	&	-99878	&	418	&	-59453	&	129	&	-11042	&	115	&	-29383	&	99	&	63	\\
1.2	&	-42799	&	179	&	-29293	&	65	&	-1949	&	42	&	-11557	&	25	&	36	\\
1.4	&	-20359	&	88	&	-15533	&	32	&	150	&	19	&	-4976	&	10	&	22	\\
1.6	&	-10300	&	58	&	-8540	&	23	&	525	&	9	&	-2285	&	5	&	14	\\
1.8	&	-5308	&	39	&	-4685	&	14	&	479	&	10	&	-1102	&	2	&	10	\\
2	&	-2675	&	17	&	-2463	&	3	&	346	&	3	&	-557.0	&	1.6	&	7	\\
2.1768	&	-1298	&	14	&	-1238	&	3	&	250	&	2	&	-310.6	&	1.2	&	6	\\
2.4	&	-277	&	11	&	-281.5	&	1.9	&	160	&	2	&	-155.3	&	0.5	&	5	\\
2.6	&	272	&	9	&	247.7	&	1.7	&	108.7	&	1.4	&	-84.6	&	0.3	&	4	\\
2.8	&	612	&	7	&	584.9	&	1.3	&	74.6	&	1.2	&	-47.70	&	0.19	&	3	\\
3	&	817	&	6	&	796.3	&	1.2	&	48.1	&	0.9	&	-27.43	&	0.12	&	3	\\
3.2	&	943	&	5	&	926.2	&	1.1	&	33.2	&	0.6	&	-16.03	&	0.08	&	2	\\
3.4	&	1020	&	4	&	1007.3	&	1.0	&	22.7	&	0.5	&	-9.50	&	0.05	&	1.8	\\
3.6	&	1059	&	4	&	1048.9	&	0.8	&	15.4	&	0.5	&	-5.68	&	0.04	&	1.6	\\
4	&	1078	&	3	&	1072.9	&	0.7	&	7.3	&	0.3	&	-2.139	&	0.016	&	1.2	\\
4.222	&	1068	&	3	&	1064.7	&	0.9	&	4.9	&	0.1	&	-1.275	&	0.009	&	1.1	\\
4.5	&	1042	&	2	&	1040.2	&	0.5	&	2.9	&	0.2	&	-0.683	&	0.006	&	0.9	\\
5	&	983.8	&	1.6	&	982.7	&	0.4	&	1.35	&	0.11	&	-0.222	&	0.003	&	0.7	\\
5.5	&	919.5	&	1.3	&	919.1	&	0.3	&	0.48	&	0.05	&	-0.076	&	0.001	&	0.6	\\
6	&	856.8	&	1.1	&	856.6	&	0.3	&	0.21	&	0.03	&	-0.029	&	0.001	&	0.5	\\    
    \hline
  \end{tabular}    
  \label{tab:CHe4_low}
\end{table*}  
\end{turnpage}

\begin{table}
  \caption{The third virial coefficient $C(T)$ and its expanded $(k=2)$
    uncertainty $U(C)$ for ${}^4$He. The last two columns
    reports the standard $(k=1)$ statistical uncertainty due to the Monte Carlo
    calculation and the combined standard $(k=1)$ uncertainty propagated from the
    pair and three-body potentials. Temperatures are in units of K and $C(T)$ in units of cm${}^6$/mol${}^2$.}
  \begin{tabular}{d|d|d||d|d}
    \multicolumn{1}{c|}{$T$} &
    \multicolumn{1}{c|}{$C$} &
    \multicolumn{1}{c||}{$U(C)$} &    
    \multicolumn{1}{c|}{$u(C_\mathrm{Boltz})$} &
    \multicolumn{1}{c|}{$u_\mathrm{pot}$} \\
    \hline
    \hline
7	&	745.5	&	0.8	&			0.2	&									0.3	\\
8	&	656.5	&	0.6	&			0.14	&									0.3	\\
9	&	586.1	&	0.5	&			0.11	&									0.2	\\
10	&	530.6	&	0.4	&			0.07	&									0.17	\\
12	&	449.5	&	0.3	&			0.08	&									0.13	\\
13.8031	&	399.4	&	0.2	&			0.06	&									0.10	\\
14	&	394.9	&	0.2	&			0.06	&									0.10	\\
15	&	374.1	&	0.2	&			0.05	&									0.09	\\
16	&	356.5	&	0.2	&			0.05	&									0.09	\\
18	&	328.33	&	0.17	&			0.04	&									0.07	\\
20	&	306.84	&	0.14	&			0.02	&									0.06	\\
22	&	290.11	&	0.13	&			0.03	&									0.06	\\
24	&	276.65	&	0.12	&			0.03	&									0.05	\\
24.5561	&	273.42	&	0.11	&			0.02	&									0.05	\\
26	&	265.66	&	0.11	&			0.03	&									0.05	\\
28	&	256.43	&	0.10	&			0.03	&									0.04	\\
30	&	248.65	&	0.09	&			0.015	&									0.04	\\
35	&	233.38	&	0.08	&			0.018	&									0.04	\\
40	&	222.00	&	0.07	&			0.010	&									0.03	\\
45	&	213.04	&	0.06	&			0.014	&									0.03	\\
50	&	205.68	&	0.05	&			0.008	&									0.03	\\
54.3584	&	200.17	&	0.05	&			0.010	&									0.02	\\
60	&	194.00	&	0.05	&			0.008	&									0.02	\\
70	&	184.79	&	0.04	&			0.006	&									0.02	\\
80	&	177.17	&	0.04	&			0.006	&									0.018	\\
83.8058	&	174.57	&	0.04	&			0.006	&									0.017	\\
90	&	170.63	&	0.03	&			0.004	&									0.016	\\
100	&	164.86	&	0.03	&			0.005	&									0.015	\\
120	&	155.07	&	0.03	&			0.003	&									0.013	\\
140	&	146.92	&	0.02	&			0.003	&									0.012	\\
160	&	139.94	&	0.02	&			0.003	&									0.011	\\
180	&	133.84	&	0.02	&			0.002	&									0.010	\\
200	&	128.44	&	0.02	&			0.003	&									0.010	\\
225	&	122.487	&	0.019	&			0.002	&									0.009	\\
234.3156	&	120.456	&	0.018	&			0.002	&									0.009	\\
250	&	117.231	&	0.018	&			0.002	&									0.009	\\
273.16	&	112.867	&	0.017	&			0.003	&									0.008	\\
300	&	108.320	&	0.017	&			0.003	&									0.008	\\
302.9146    &   107.860 &   0.016   &           0.001   &                                   0.008   \\
325 &   104.495 &   0.018   &    0.004   &                                   0.008   \\
350 &   101.003 &   0.015   &    0.001   &                                   0.007   \\
375 &   97.796  &   0.016   &    0.003   &                                   0.007   \\
400 &   94.840  &   0.014   &    0.002   &                                   0.007   \\
429.7485	&	91.602	&	0.013	&			0.0011	&									0.007	\\
450	&	89.550	&	0.013	&			0.0014	&									0.007	\\
500	&	84.944	&	0.013	&			0.0018	&									0.006	\\
600	&	77.274	&	0.012	&			0.0013	&									0.006	\\
700	&	71.097	&	0.011	&			0.0010	&									0.005	\\
800	&	65.983	&	0.011	&			0.0011	&									0.005	\\
1000	&	57.939	&	0.010	&			0.0012	&									0.005	\\
1200	&	51.848	&	0.009	&			0.0007	&									0.005	\\
1500	&	44.974	&	0.009	&			0.0007	&									0.004	\\
2000	&	37.043	&	0.008	&			0.0007	&									0.004	\\
2500	&	31.596	&	0.008	&			0.0006	&									0.004	\\
3000	&	27.579	&	0.008	&			0.0009	&									0.004	\\    
    \hline
  \end{tabular}    
  \label{tab:CHe4_high}
\end{table}

In order to facilitate comparisons with other results in the literature, we developed a correlation
for the values of a generic virial coefficient $F(T)$ and its expanded uncertainty $U(F)$ in the
form of
\begin{eqnarray}
  F(T) &=& \sum_{k=1}^n \frac{a_k}{(T/T_0)^{b_{k}}}, \label{eq:correlation} \\
  U(F) &=& a \exp\left( \frac{b}{(T/T_0)^c} \right) \label{eq:UCcorrelation},
\end{eqnarray}
which smoothly interpolates the Monte Carlo data for $F = C$, passing within the expanded $(k=2)$
uncertainty $U(F)$ at all the temperatures in the range $2~\mathrm{K} \leq T \leq 3000$~K. The limitation in the
temperature range of the correlation is because the rapid decrease of $C(T)$ below
$T=2$~K prevented us from finding a satisfactory set of parameters for Eq.~(\ref{eq:correlation}).
The parameters of Eq.~(\ref{eq:correlation}) for helium isotopologues are given in
Table~\ref{tab:correlation}, while those for Eq.~(\ref{eq:UCcorrelation}) are given in
Table~\ref{tab:UCcorrelation}. We also report in Table~\ref{tab:correlation} fitting parameters,
assuming the same form as Eq.~(\ref{eq:correlation}), obtained for $F = B$ using the data
computed in Ref.~\onlinecite{u2_2020}. The correlations for $B(T)$ and $C(T)$ enable the calculation
of $\RTg$ (see Eqs.~(\ref{eq:ba})--(\ref{eq:RTg})), and we have used these values as a cross check of
the values obtained using the PIMC and SPM approaches.

\begin{table*}
  \caption{Parameters of the correlation in the right-hand side of Eq.~(\ref{eq:correlation}) for
    the second and third virial coefficient of ${}^4$He and ${}^3$He. For both isotopes, the
    correlations have been fitted with data in the range $2 - 3000$~K and are therefore not reliable
    outside this range. $T_0$ is given in K, the parameters $b_i$ are dimensionless, and the
    parameters $a_i$ have units of cm${}^3$/mol for $B$ and cm${}^6$/mol${}^2$ for $C$.}
    \begin{tabular}{c|c|c||c|c}
      Parameter & $B({}^4$He) & $C({}^4$He) & $B({}^3$He) & $C({}^3$He) \\
      \hline
      \hline
    $n$   &   6 &   8 &   6 &   6 \\
    $T_0$ & 100 & 100 & 100 & 100 \\
    \hline
    $a_1$ & $-2.313629 \times 10^{-7}$ & $-8.677343\times 10^{-9}$ & $-4.753277 \times 10^{-10}$ & $-7.844580 \times 10^{-4}$ \\
    $a_2$ & $-2.600944 \times 10^{-1}$ & $-1.772954\times 10^{1} $ & $ 1.976692 \times 10^{ -8}$ & $-3.968348 \times 10^{2}$ \\
    $a_3$ & $-5.991829$                & $ 1.501863\times 10^{2} $ & $-1.743840 \times 10^{ -1}$ & $ 1.242512 \times 10^{3}$ \\
    $a_4$ & $-3.191519 \times 10^{ 2}$ & $-5.865689\times 10^{2} $ & $-1.431270 \times 10^{  1}$ & $-2.009221 \times 10^{3}$ \\
    $a_5$ & $ 6.694613 \times 10^{ 2}$ & $ 1.086422\times 10^{3} $ & $ 6.478292 \times 10^{  1}$ & $ 2.054704 \times 10^{3}$ \\
    $a_6$ & $-3.323825 \times 10^{ 2}$ & $-1.671338\times 10^{3} $ & $-3.825692 \times 10^{  1}$ & $-7.205123 \times 10^{2}$ \\
    $a_7$ & ---                        & $ 1.863898\times 10^{3} $ & --- & --- \\
    $a_8$ & ---                        & $-6.600046\times 10^{2} $ & --- & --- \\
    \hline
    $b_1$ & $4.327481$ & $6.538462$ & 6        & 3.740741 \\
    $b_2$ & $1.406369$ & $2.142857$ & 5.162791 & 1.272720 \\
    $b_3$ & $0.801565$ & $1.823529$ & 1.363636 & 1.142857 \\
    $b_4$ & $0.301418$ & $1.526316$ & 0.641026 & 0.9      \\
    $b_5$ & $0.260488$ & $1.277778$ & 0.2      & 0.64     \\
    $b_6$ & $0.225976$ & $0.923077$ & 0.135135 & 0.5      \\
    $b_7$ & ---        & $0.642857$ & ---       & --- \\
    $b_8$ & ---        & $0.5     $ & ---       & --- \\
    \hline
  \end{tabular}
  \label{tab:correlation}  
\end{table*}

\begin{table*}
  \caption{Parameters of the correlation in Eq.~(\ref{eq:UCcorrelation}) for
    the expanded uncertainty of the third virial coefficient of ${}^4$He and ${}^3$He.
    $T_0$ is given in K, $a$ in cm${}^6$/mol${}^2$, $b$ and $c$ are dimensionless. For both isotopes, the correlations have been fitted with data
    in the range $2 - 3000$~K and are therefore not reliable
    outside this range.}
    \begin{tabular}{c|c|c}
      Parameter & ${}^4$He & ${}^3$He \\
      \hline
      \hline
      $T_0$ & 100 & 100 \\
    \hline
    $a$ & $6.506 \times 10^{-3}$ & $5.07 \times 10^{-3}$ \\
    $b$ & 1.59486               & 1.80741 \\
    $c$ & 0.422092              & 0.379443 \\    
    \hline
  \end{tabular}
  \label{tab:UCcorrelation}  
\end{table*}

\begin{figure}
  \center\includegraphics[width=0.8\linewidth]{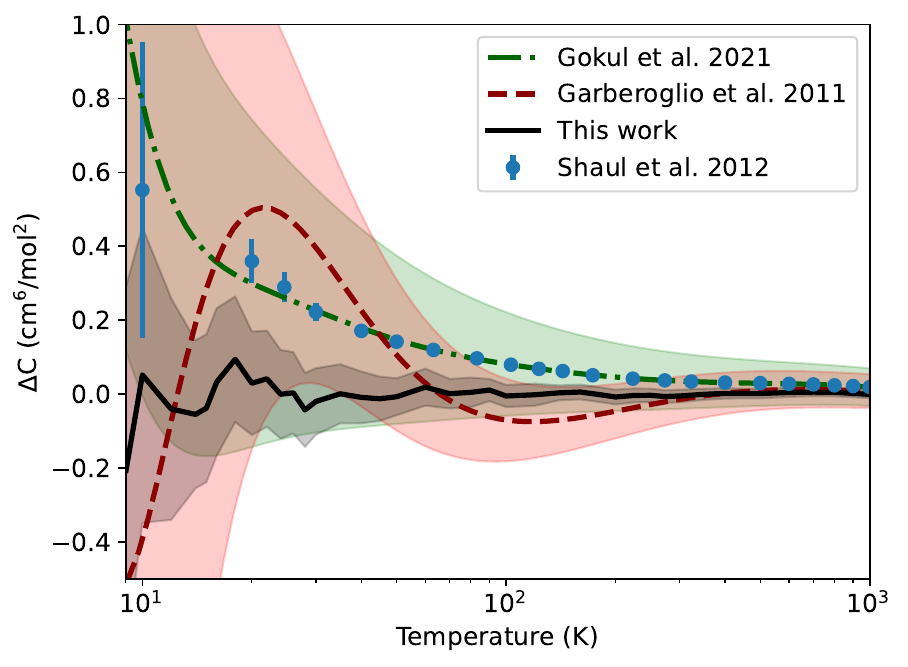}    
  \caption{Comparison of literature results for $C(T)$ in the case of
    ${}^4$He. We use as a baseline the correlation of 
    Eq.~(\ref{eq:correlation}). The black line and gray shading show the results and the expanded
    $(k=2)$ uncertainty of the present calculations. The red area covers the
    results of our previous calculation in
    2011.~\cite{Garberoglio2011} The dashed green line and the green area
    reports the results from the correlation developed by Gokul {\em et
      al.}~\cite{Gokul21} The blue dots are the results by Shaul {\em et
    al.}~\cite{Shaul2012} In this latter case, the uncertainty is the
    expanded uncertainty of the Monte Carlo calculation.}  
\label{fig:He4_comp}
\end{figure}

The updated values of $C(T)$ computed in this work are compared with other
literature results in Fig.~\ref{fig:He4_comp}.
In general, the values obtained in the present work are compatible, within
mutual uncertainties, with the other calculations. However, due to the use of
updated two- and three-body potentials, the uncertainty is much smaller.
Compared with the values of $C(T)$ obtained by the Kofke
group,~\cite{Shaul2012,Gokul21} the present data show a downward
shift. This is consistent with the finding that, due to relativistic
effects, the new three-body potential is generally more attractive than the
non-relativistic one used in all previous calculations.~\cite{u3_2023}

When compared to our previous results,~\cite{Garberoglio2011} this rigid shift is not
apparent, although the two sets of values are mutually compatible. This
might indicate incomplete convergence of some parameters (cutoff, number
of beads) in our previous calculation.

\begin{figure}
  \center\includegraphics[width=0.8\linewidth]{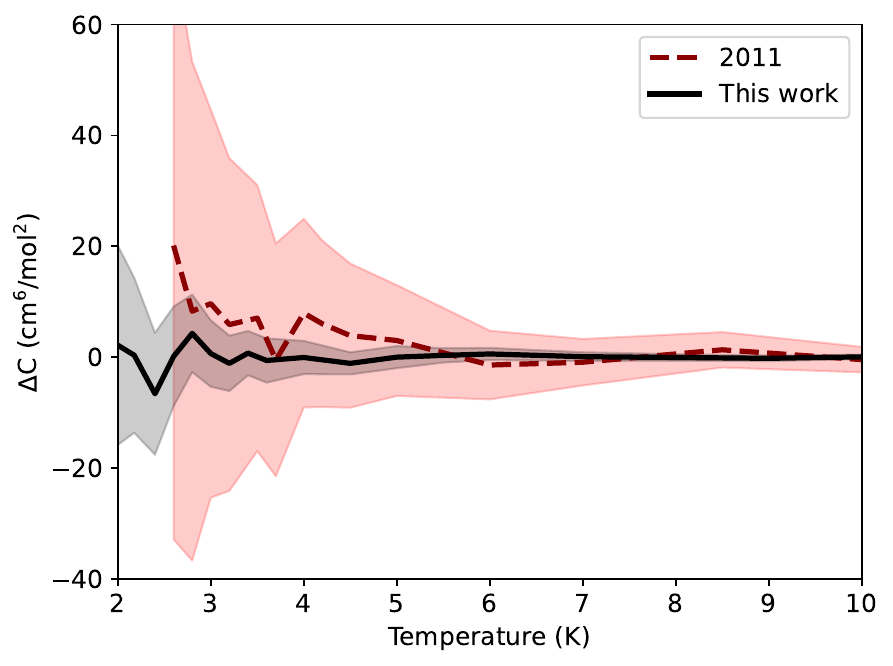}    
  \caption{Comparison with our previous results for $C(T)$ at low
    temperatures for ${}^4$He.  We use as a baseline the correlation of
    Eq.~(\ref{eq:correlation}). The black line and gray shading show the
    results and the expanded $(k=2)$ uncertainty of the present
    calculations.  The results of our previous calculation in
    2011~\cite{Garberoglio2011,Garberoglio2011a} are shown as a dashed red
    curve with light red shading for the expanded uncertainty.}
\label{fig:He4_comp_low}
\end{figure}

In the case of the lowest temperatures, where exchange effects are
significant, our new values are in very good agreement with results reported
earlier.~\cite{Garberoglio2011,Garberoglio2011a} The data, shown in
Fig.~\ref{fig:He4_comp_low}, agree within mutual uncertainties in
the whole temperature range where the calculations overlap. In this case,
the expected downward shift of $C(T)$ due to the more attractive three-body
potential~\cite{u3_2023} is much more evident that at the highest temperatures.
This figure also shows the reduction in the uncertainty due to the much
more accurate pair and three-body potentials used in this work.

\subsubsection{Third acoustic virial coefficient, $\RTg$}

\begin{figure}
  \center\includegraphics[width=0.8\linewidth]{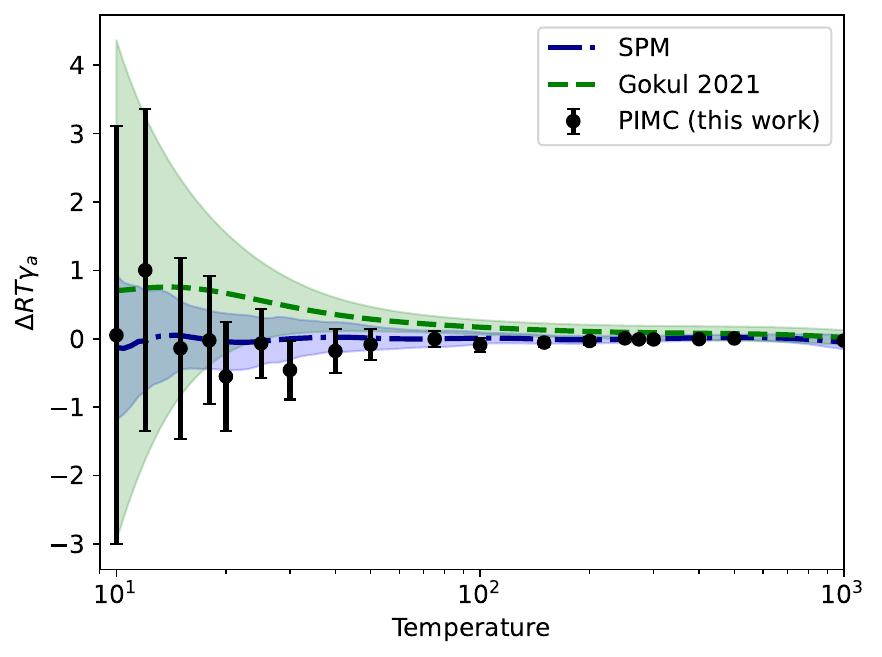}    
  \caption{Comparison of calculations of the third acoustic virial
    coefficient, $\RTg$, for ${}^4$He. The baseline is the value obtained
    using the correlations of Eq.~(\ref{eq:correlation}) for $B(T)$ and
    $C(T)$.  The dot-dashed line reports the SPM evaluation of $\RTg$
    obtained from the $C(T)$ curve computed in this work and the $B(T)$
    data from Ref.~\onlinecite{u2_2020}. The blue shaded area reports the
    expanded uncertainty of $\RTg$ propagated using SPM.  The black points
    are the path-integral results obtained in this work.  The green curve
    reports the values obtained by Gokul {\em et al.}~\cite{Gokul21}
    together with their estimated expanded uncertainty.}
\label{fig:RTg_He4}  
\end{figure}

Our results for the third acoustic virial coefficient for ${}^4$He are
reported in Table~\ref{tab:RTg_He4} and illustrated in
Fig.~\ref{fig:RTg_He4}.
The values that we obtain are generally compatible with those calculated by
Gokul {\em et al.},~\cite{Gokul21} although also in this case a systematic
downward shift -- which we ascribe to the improved three-body potential -- is
more evident.

Figure~\ref{fig:RTg_He4} also reports the results of the SPM approach, obtained from the
path-integral values of $C(T)$ computed in this work and the $B(T)$ values
of Ref.~\onlinecite{u2_2020}.
As mentioned above, computer time limitations prevented us from reducing the
statistical uncertainty of $\RTg$ to a value comparable to the propagated
uncertainty from the potential, in contrast to what we were able to do
in the case of $C(T)$, at temperatures $T \leq 75$~K.
At higher temperatures, the path-integral results are in very good
agreement with those obtained by the SPM approach, although we notice that
the SPM uncertainty is larger than the path-integral one for $T > 500$~K.
In the low-temperature regime, the path-integral results follow closely the
SPM values, whose uncertainty is, however, smaller.
We did not develop our calculation methods for $\RTg$ to include exchange effects; hence
the SPM approach is presently the only way to compute the third acoustic virial coefficient of helium
isotopologues below $\approx 10$~K and we suggest its use for $1$~K~$\leq T \leq 75$~K. A
table of SPM values for ${}^4$He in the temperature range $1 - 1000$~K is provided as Supplementary Material.

\begin{table}
  \caption{Values of $\RTg$ for ${}^4$He calculated in this work using
    PIMC or the SPM method, together with their
    expanded $(k=2)$ uncertainties. Temperatures are in units of K and
  $\RTg$ in units of cm${}^6$/mol${}^2$.}
  \begin{tabular}{d|d|d|d|d}
    \multicolumn{1}{c|}{$T$} &
    \multicolumn{1}{c|}{$\RTg$} &
    \multicolumn{1}{c|}{$U(\RTg)$} &
    \multicolumn{1}{c|}{$\RTg$} &
    \multicolumn{1}{c}{$U(\RTg)$} \\
    \hline
    \multicolumn{1}{c|}{} &
    \multicolumn{2}{c|}{PIMC} &
    \multicolumn{2}{c}{SPM} \\
    \hline
    \hline
     2	&	\multicolumn{1}{c|}{--}	& \multicolumn{1}{c|}{--}	&	-29358	&	125	\\    
     4	&	\multicolumn{1}{c|}{--}	& \multicolumn{1}{c|}{--}	&	-5246	&	9	\\    
     6	&	\multicolumn{1}{c|}{--}	& \multicolumn{1}{c|}{--}	&	-680	&	4	\\    
     8 	&	\multicolumn{1}{c|}{--}	& \multicolumn{1}{c|}{--}	&	457.6	&	1.9	\\    
    10	&	806	&	3	&	806.1	&	1.1	\\
    12	&	906	&	2	&	905.2	&	0.7	\\
15	&	900.6	&	1.3	&	900.8	&	0.5	\\
18	&	841.5	&	0.9	&	841.5	&	0.4	\\
20	&	794.7	&	0.8	&	795.2	&	0.4	\\
25	&	684.5	&	0.5	&	684.54	&	0.34	\\
30	&	591.0	&	0.4	&	591.42	&	0.30	\\
40	&	453.6	&	0.3	&	453.76	&	0.22	\\
50	&	360.2	&	0.2	&	360.25	&	0.18	\\
75	&	224.07	&	0.11	&	224.07	&	0.12	\\
100	&	151.96	&	0.11	&	152.06	&	0.09	\\
150	&	78.95	&	0.07	&	78.99	&	0.06	\\
200	&	43.10	&	0.06	&	43.12	&	0.06	\\
250	&	22.40	&	0.05	&	22.38	&	0.05	\\
273.16	&	15.58	&	0.05	&	15.59	&	0.05	\\
300	&	9.17	&	0.05	&	9.18	&	0.04	\\
400	&	-6.16	&	0.04	&	-6.14	&	0.04	\\
500	&	-14.35	&	0.04	&	-14.33	&	0.04	\\
1000	&	-26.10	&	0.03	&	-26.12	&	0.11	\\
1500	&	-27.09	&	0.02	&	\multicolumn{1}{c|}{--}	& \multicolumn{1}{c}{--}	\\
2000	&	-26.38	&	0.02	&	\multicolumn{1}{c|}{--}	& \multicolumn{1}{c}{--}	\\
    \hline
  \end{tabular}
  \label{tab:RTg_He4}
\end{table}

\subsubsection{Comparison with experiment}
Extensive comparisons with experimental data for $C(T)$ and $RT\gamma_\mathrm{a}(T)$ for $^4$He were
given in previous work.~\cite{Garberoglio2009b,Garberoglio2011} Already at that time, the uncertainties of
calculated values were much smaller than those from experiment. We therefore limit our comparisons
for $C(T)$ to cryogenic temperatures and to a few high-accuracy experimental sources near room
temperature.

\begin{figure}
\includegraphics[width=0.9\linewidth]{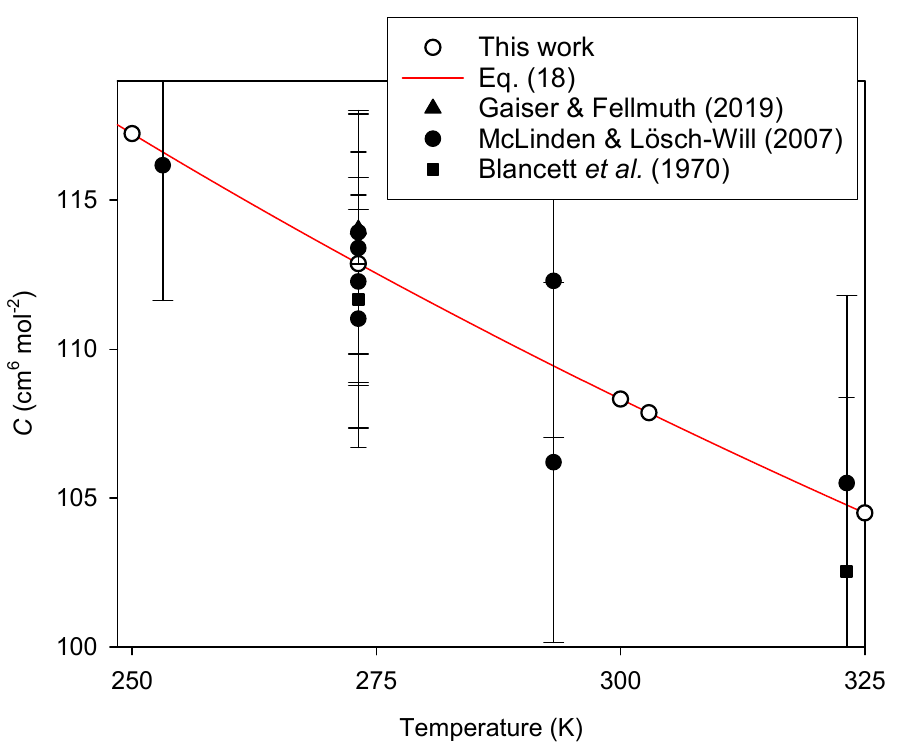}  
  \caption{Comparison of calculated $C(T)$ for $^4$He with experimental
    values at near-ambient temperatures. Error bars on experimental points
    represent expanded uncertainties with coverage factor $k$ = 2; expanded
    uncertainties for this work (see Table~\ref{tab:CHe4_high}) are much
    smaller than the symbols.}
\label{fig:roomT}
\end{figure}

Figure \ref{fig:roomT} shows $C(T)$ near room temperature as points from Table~\ref{tab:CHe4_high}
and as given by the fitting equation~(\ref{eq:correlation}).  Experimental values of $C$ from three
sources are plotted.\cite{Blancett_1970,McLinden_2007,Gaiser_2019} For the datum of Gaiser and
Fellmuth\cite{Gaiser_2019} from dielectric-constant gas thermometry at 273.16~K, the quantity
reported was a combination of density and dielectric virial coefficients; this was converted to $C$
using the best calculated values for the second\cite{Garberoglio_2020} and third~\cite{Ceps_2024}
dielectric virial coefficients.  Our results are consistent with these state-of-the-art experiments,
but our uncertainties (which are much smaller than the size of the symbols) are smaller by at least
two orders of magnitude.

\begin{figure}
  \includegraphics[width=0.9\linewidth]{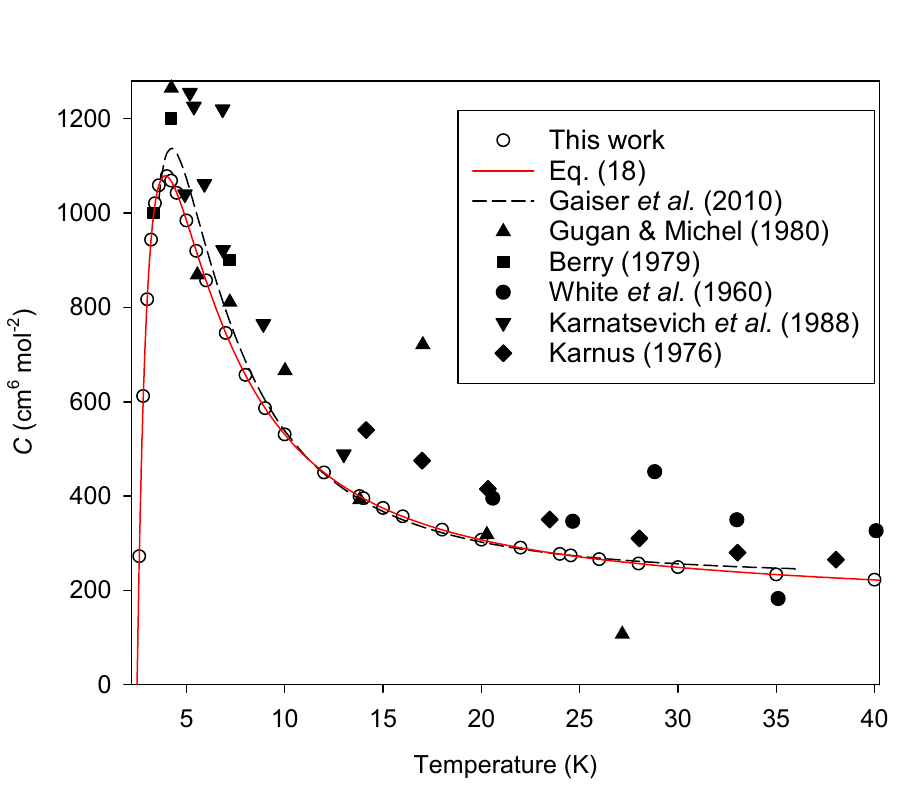}  
  \caption{Comparison of calculated $C(T)$ for $^4$He with experimental
    values at low temperatures. Error bars on experimental points are not
    shown for clarity (see text); expanded uncertainties for this work (see
    Table~\ref{tab:CHe4_high}) are smaller than the symbols.}
\label{fig:lowT}
\end{figure}

Figure \ref{fig:lowT} shows a similar comparison for $C(T)$ below 40~K, where the available
experimental
sources\cite{White_1960,Karnus_1976,Berry_1979,Gugan_1980,Karnatsevich_1988,Gaiser_2010} are
somewhat scattered. For clarity, we do not show error bars for the experimental sources; in some
cases they were not reported, while in others they were on the order of several hundred
cm$^6$~mol$^{-2}$.  The uncertainty of our results is smaller than the size of the symbols; this can
be compared with Fig.~2 of Ref.~\onlinecite{Garberoglio2011} where error bars for calculated
results were visible below 5~K.  Our results follow the general trend of the experimental data, but
again have much smaller uncertainties. 

\begin{figure}
\includegraphics[width=0.9\linewidth]{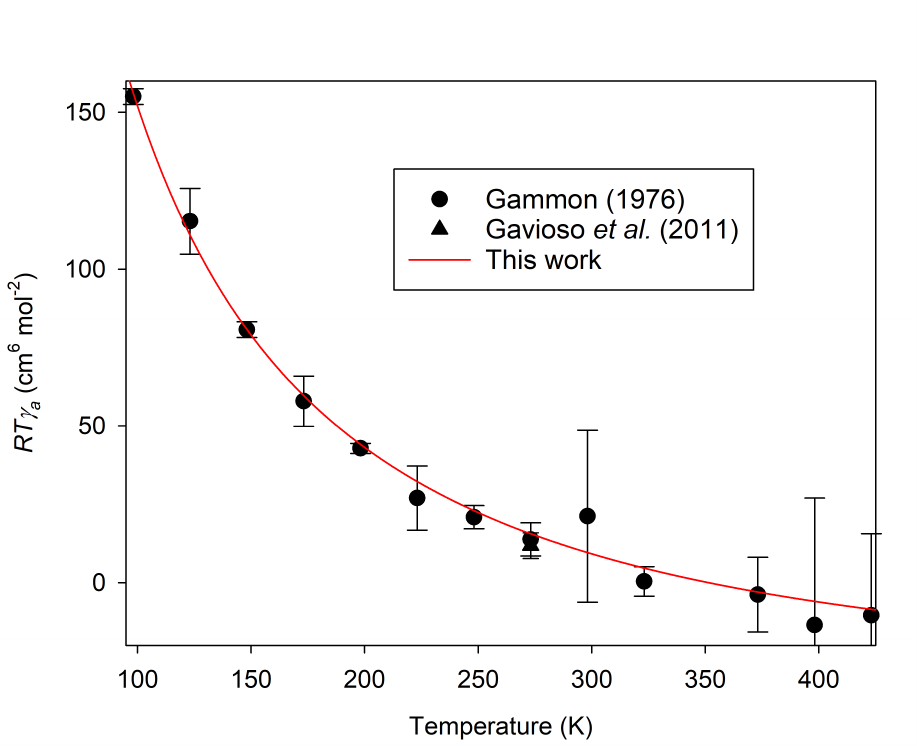}  
  \caption{Comparison of calculated $RT\gamma_\mathrm{a}(T)$ for $^4$He
    with experimental values. Error bars on experimental points represent
    expanded uncertainties with coverage factor $k$ = 2; expanded
    uncertainties for this work (see Table~\ref{tab:RTg_He4}) are no larger
    than the width of the curve.}
\label{fig:Acoustic}
\end{figure}

The situation for the third acoustic virial coefficient is similar; agreement with available
experimental data but with much smaller uncertainty.  Figure \ref{fig:Acoustic} shows our calculations
for $^4$He compared to the values of $RT\gamma_\mathrm{a}(T)$ derived in
Ref.~\onlinecite{Garberoglio2011} from the sound-speed data of Gammon,\cite{Gammon_1976} along with
a single value at 273.16~K reported by Gavioso \textit{et al.}\cite{Gavioso_2011} The agreement with
experiment is good, but our uncertainties are smaller by at least one order of magnitude.

\begin{figure}
  \includegraphics[width=0.9\linewidth]{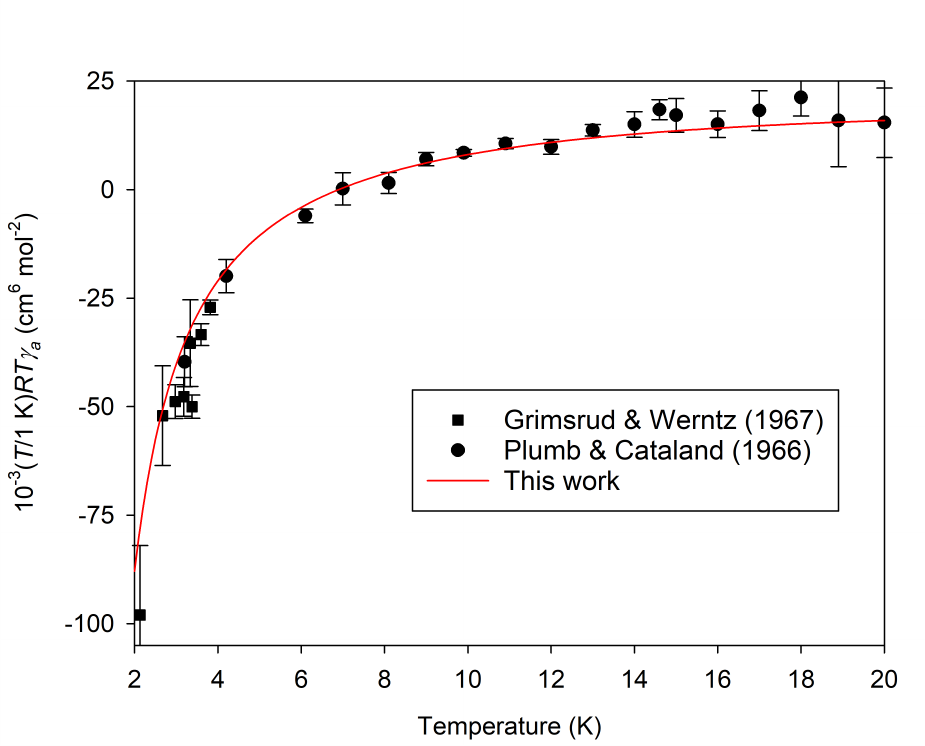}  
  \caption{Comparison of calculated $RT\gamma_\mathrm{a}(T)$ for $^4$He
    with experimental values at low temperatures. Error bars on
    experimental points represent expanded uncertainties with coverage
    factor $k$ = 2; expanded uncertainties for this work (see
    Table~\ref{tab:RTg_He4}) are no larger than the width of the curve.}
\label{fig:Acoustic-low}
\end{figure}

Figure \ref{fig:Acoustic-low} provides a similar comparison at temperatures below 20~K, where there
are two experimental data sources.\cite{Grimsrud_1967,Plumb_1966} Again we use values of
$RT\gamma_\mathrm{a}(T)$ derived from these sound-speed data in
Ref.~\onlinecite{Garberoglio2011}. In this case, we have rescaled the ordinate for clarity and
plotted the quantity 10$^{-3}(T/1~\mathrm{K})RT\gamma_\mathrm{a}$. Once again, our calculated
results are consistent with experiment but have much smaller uncertainties.

\subsection{Helium-3}

\subsubsection{Third virial coefficient, $C(T)$}

\begin{figure}
  \center\includegraphics[width=0.8\linewidth]{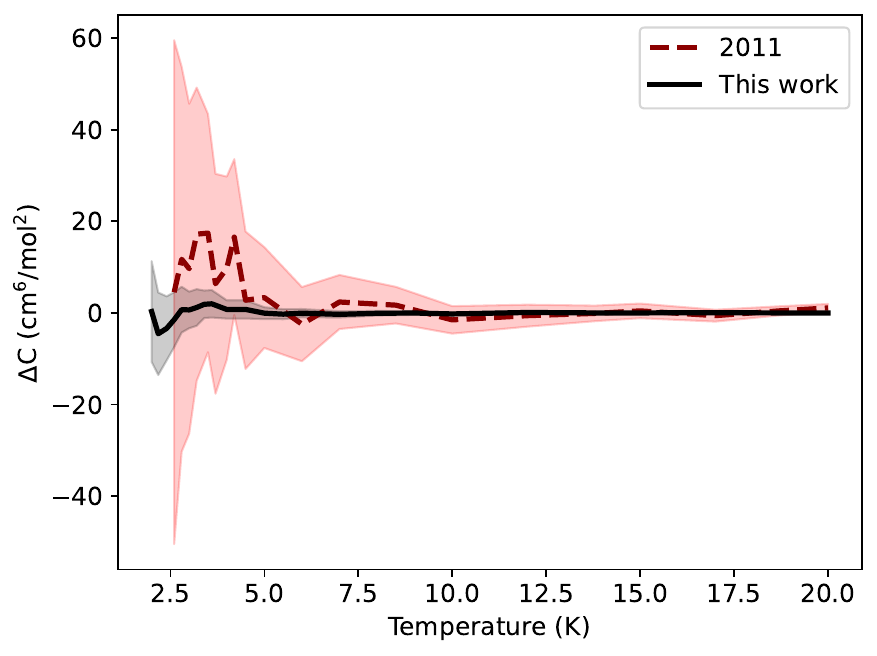}  
  \center\includegraphics[width=0.8\linewidth]{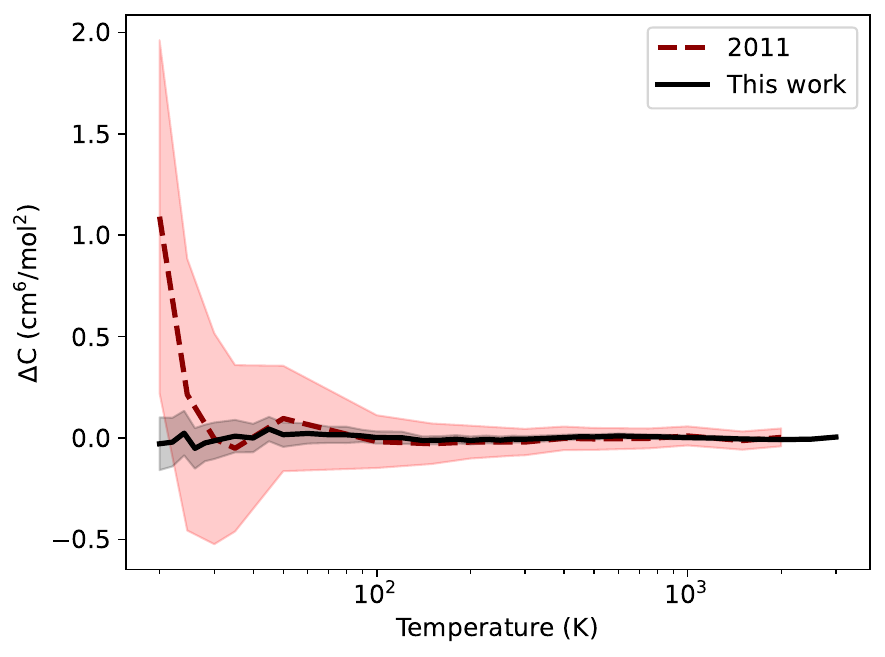}      
  \caption{Comparison between the values of $C(T)$ for ${}^3$He calculated
    in this work (black solid line) and those available in the
    literature~\cite{Garberoglio2011} (red dashed line).  We use as a
    baseline the correlation of Eq.~(\ref{eq:correlation}).  The shaded
    areas show the expanded $(k=2)$ uncertainty of the calculation that
    includes the statistical uncertainty and the uncertainty propagated
    from the pair and three-body potentials.  Upper panel: lower
    temperatures. Lower panel: higher temperatures.}
  \label{fig:CHe3}
\end{figure}

The situation for ${}^3$He closely parallels that for ${}^4$He. The improved pair and three-body
potentials result in a roughly $5$ times smaller uncertainty of $C(T)$ compared with our previously
published results.~\cite{Garberoglio2011} The values of $C(T)$ computed in this work are reported in
Tables~\ref{tab:CHe3_low} and \ref{tab:CHe3_high} for the low and high temperature case,
respectively. A graphical comparison with previous work is shown in Fig.~\ref{fig:CHe3}, where one
can see that our new values are in excellent agreement with the older results. Analogously to
${}^4$He, we extended our calculations down to $0.5$~K, rigorously including fermionic exchange
effects.~\cite{Garberoglio2011a,Garberoglio2011}

\begin{table*}
  \caption{The third virial coefficient $C(T)$ and its expanded $(k=2)$
    uncertainty $U(C)$ for ${}^3$He at low temperatures. The other columns
    report the contributions due to Boltzmann statistics
    ($C_\mathrm{Boltz}$) and the odd ($C_\mathrm{odd}$) and even exchange
    ($C_\mathrm{even}$), together with their standard $(k=1)$ statistical
    uncertainties due to the Monte Carlo calculation. The last column
    reports the combined standard $(k=1)$ uncertainty propagated from the
    pair and three-body potentials. Temperatures are in units of K and $C(T)$ in units of cm${}^6$/mol${}^2$.}
  \begin{tabular}{d|d|d||d|d|d|d|d|d|d}
    \multicolumn{1}{c|}{$T$} &
    \multicolumn{1}{c|}{$C$} &
    \multicolumn{1}{c||}{$U(C)$} &
    \multicolumn{1}{c|}{$C_\mathrm{Boltz}$} &
    \multicolumn{1}{c|}{$u(C_\mathrm{Boltz})$} &
    \multicolumn{1}{c|}{$C_\mathrm{odd}$} &
    \multicolumn{1}{c|}{$u(C_\mathrm{odd})$} &
    \multicolumn{1}{c|}{$C_\mathrm{even}$} &
    \multicolumn{1}{c|}{$u(C_\mathrm{even})$} &
    \multicolumn{1}{c}{$u_\mathrm{pot}$} \\
    \hline
    \hline
0.5	&	-98152	&	2883	&	-71905	&	982	&	47368	&	912	&	-73614	&	451	&	282	\\
0.6	&	-60497	&	1381	&	-38219	&	474	&	18060	&	452	&	-40338	&	155	&	157	\\
0.8	&	-24901	&	512	&	-11514	&	206	&	1621	&	134	&	-15009	&	35	&	62	\\
1	&	-11367	&	179	&	-3608	&	64	&	-1215	&	53	&	-6544	&	10	&	32	\\
1.2	&	-4926	&	90	&	-329	&	32	&	-1430	&	25	&	-3166	&	7	&	19	\\
1.4	&	-1694	&	51	&	1068	&	18	&	-1140	&	13	&	-1623	&	4	&	12	\\
1.6	&	3	&	32	&	1673	&	11	&	-794	&	7	&	-875.7	&	1.8	&	8	\\
1.8	&	840	&	22	&	1908	&	8	&	-576	&	4	&	-491.9	&	1.2	&	6	\\
2	&	1285	&	11	&	1962	&	2	&	-394.9	&	1.7	&	-281.9	&	0.6	&	5	\\
2.1768	&	1505	&	9	&	1968.6	&	1.7	&	-286.5	&	1.6	&	-177.5	&	0.4	&	4	\\
2.4	&	1627	&	7	&	1914.5	&	1.3	&	-187.0	&	1.3	&	-100.8	&	0.3	&	3	\\
2.6	&	1653	&	6	&	1846.9	&	1.1	&	-132.7	&	1.0	&	-61.51	&	0.18	&	3	\\
2.8	&	1637	&	5	&	1768.2	&	0.9	&	-92.8	&	0.8	&	-38.63	&	0.10	&	2	\\
3	&	1596	&	4	&	1688.0	&	0.8	&	-67.5	&	0.7	&	-24.59	&	0.07	&	1.8	\\
3.2	&	1544	&	4	&	1607.8	&	0.7	&	-48.0	&	0.6	&	-15.91	&	0.05	&	1.5	\\
3.4	&	1487	&	3	&	1531.0	&	0.7	&	-33.6	&	0.5	&	-10.33	&	0.03	&	1.3	\\
3.6	&	1428	&	3	&	1459.0	&	0.6	&	-24.5	&	0.3	&	-6.81	&	0.02	&	1.2	\\
4	&	1312	&	2	&	1329.1	&	0.5	&	-13.7	&	0.2	&	-3.083	&	0.010	&	0.9	\\
4.5	&	1183.3	&	1.6	&	1190.8	&	0.4	&	-6.35	&	0.15	&	-1.190	&	0.006	&	0.7	\\
5	&	1071.6	&	1.3	&	1075.1	&	0.3	&	-3.03	&	0.09	&	-0.488	&	0.002	&	0.6	\\
5.5	&	977.7	&	1.1	&	979.5	&	0.3	&	-1.57	&	0.05	&	-0.204	&	0.001	&	0.5	\\
6	&	898.6	&	0.9	&	899.3	&	0.2	&	-0.59	&	0.08	&	-0.090	&	0.001	&	0.4	\\
7	&	773.2	&	0.7	&	773.45	&	0.16	&	-0.23	&	0.04	&	-0.019	&	0.000	&	0.3	\\
8	&	680.7	&	0.5	&	680.76	&	0.12	&	-0.06	&	0.02	&	-0.004	&	0.000	&	0.2	\\    
    \hline
  \end{tabular}    
  \label{tab:CHe3_low}
\end{table*}

\begin{table}  
  \caption{The third virial coefficient $C(T)$ and its expanded $(k=2)$
    uncertainty $U(C)$ for ${}^3$He. The last two columns
    reports the standard $(k=1)$ statistical uncertainty due to the Monte Carlo
    calculation and the combined standard $(k=1)$ uncertainty propagated from the
    pair and three-body potentials. Temperatures are in units of K and $C(T)$ in units of cm${}^6$/mol${}^2$.}
  \begin{tabular}{d|d|d||d|d}
    \multicolumn{1}{c|}{$T$} &
    \multicolumn{1}{c|}{$C$} &
    \multicolumn{1}{c||}{$U(C)$} &    
    \multicolumn{1}{c|}{$u(C_\mathrm{Boltz})$} &
    \multicolumn{1}{c|}{$u_\mathrm{pot}$} \\
    \hline
    \hline
9	&	610.2	&	0.4	&			0.10	&									0.18	\\
10	&	555.0	&	0.3	&			0.09	&									0.15	\\
12	&	476.0	&	0.3	&			0.07	&									0.11	\\
14	&	422.1	&	0.2	&			0.05	&									0.09	\\
15	&	401.3	&	0.2	&			0.05	&									0.08	\\
16	&	383.62	&	0.18	&			0.05	&									0.08	\\
18	&	354.78	&	0.15	&			0.04	&									0.07	\\
20	&	332.41	&	0.13	&			0.03	&									0.06	\\
22	&	314.63	&	0.12	&			0.02	&									0.05	\\
24	&	300.15	&	0.11	&			0.03	&									0.05	\\
26	&	287.97	&	0.10	&			0.03	&									0.05	\\
28	&	277.73	&	0.09	&			0.02	&									0.04	\\
30	&	268.90	&	0.09	&			0.019	&									0.04	\\
35	&	251.28	&	0.08	&			0.016	&									0.03	\\
40	&	237.90	&	0.07	&			0.015	&									0.03	\\
45	&	227.30	&	0.06	&			0.015	&									0.03	\\
50	&	218.48	&	0.06	&			0.012	&									0.02	\\
60	&	204.53	&	0.05	&			0.010	&									0.02	\\
70	&	193.66	&	0.04	&			0.010	&									0.019	\\
80	&	184.76	&	0.04	&			0.007	&									0.017	\\
90	&	177.21	&	0.03	&			0.007	&									0.016	\\
100	&	170.65	&	0.03	&			0.006	&									0.015	\\
120	&	159.67	&	0.03	&			0.004	&									0.013	\\
140	&	150.67	&	0.02	&			0.004	&									0.012	\\
160	&	143.08	&	0.02	&			0.003	&									0.011	\\
180	&	136.53	&	0.02	&			0.004	&									0.010	\\
200	&	130.77	&	0.02	&			0.002	&									0.010	\\
225	&	124.468	&	0.019	&			0.002	&									0.009	\\
250	&	118.936	&	0.018	&			0.002	&									0.009	\\
273.16	&	114.379	&	0.017	&			0.002	&									0.008	\\
300	&	109.641	&	0.016	&			0.002	&									0.008	\\
350	&	102.060	&	0.015	&			0.002	&									0.007	\\
400	&	95.711	&	0.014	&			0.001	&									0.007	\\
450	&	90.286	&	0.013	&			0.001	&									0.006	\\
500	&	85.573	&	0.013	&			0.001	&									0.006	\\
600	&	77.755	&	0.012	&			0.001	&									0.006	\\
700	&	71.475	&	0.011	&			0.001	&									0.006	\\
800	&	66.291	&	0.011	&			0.001	&									0.005	\\
1000	&	58.160	&	0.010	&			0.001	&									0.005	\\
1200	&	52.014	&	0.010	&			0.001	&									0.005	\\
1500	&	45.091	&	0.009	&			0.001	&									0.004	\\
2000	&	37.119	&	0.008	&			0.001	&									0.004	\\
2500	&	31.646	&	0.008	&			0.001	&									0.004	\\
3000	&	27.617	&	0.007	&			0.001	&									0.004	\\    
    \hline
  \end{tabular}    
  \label{tab:CHe3_high}
\end{table}

\subsubsection{Third acoustic virial coefficient, $\RTg$}

To the best of our knowledge, no values of $\RTg$ for ${}^3$He have
appeared in the literature. We report our calculated values in Table~\ref{tab:RTg_He3}. Also in this
case, we used SPM to propagate the uncertainty, as well as performing the calculation directly. In
this case, the agreement between these two methods is very good, except at temperatures $T \gtrsim
500$~K for which, as already noted in Sec.~\ref{sec:beta_SPM}, the SPM approach overestimates the
uncertainty.
Also for this isotope, SPM is the only way we can provide values of $\RTg$ and its uncertainty at
temperatures where exchange effects are significant. For these reasons, we recommend the use of
SPM-derived values of $\RTg$ and its uncertainty, which are reported as Supplementary Material.

\begin{figure}
  \center\includegraphics[width=0.8\linewidth]{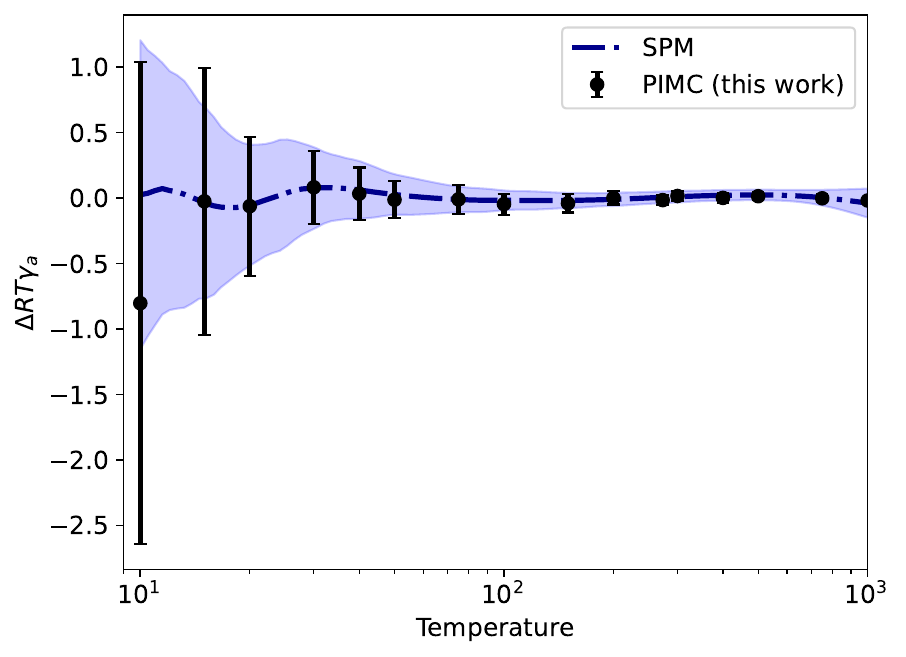}   
  \caption{Comparison of calculations of the third acoustic virial
    coefficient, $\RTg$, for ${}^3$He. The baseline is the value obtained
    using the correlation of Eq.~(\ref{eq:correlation}) for $B(T)$ and
    $C(T)$.  The dot-dashed line reports the SPM evaluation of $\RTg$
    obtained from the $C(T)$ curve computed in this work, and the $B(T)$
    data from Ref.~\onlinecite{u2_2020}. The blue shaded area shows the
    expanded uncertainty of $\RTg$ propagated using SPM.  The black points
    are the path-integral results obtained in this work.  }
\end{figure}

\begin{table}
  \caption{Values of $\RTg$ for ${}^3$He calculated in this work using
    PIMC or the SPM method, together with their
    expanded $(k=2)$ uncertainties. Temperatures are in units of K and
  $\RTg$ in units of cm${}^6$/mol${}^2$.}
  \begin{tabular}{d|d|d|d|d}
    \multicolumn{1}{c|}{$T$} &
    \multicolumn{1}{c|}{$\RTg$} &
    \multicolumn{1}{c|}{$U(\RTg)$} &
    \multicolumn{1}{c|}{$\RTg$} &
    \multicolumn{1}{c}{$U(\RTg)$} \\
    \hline
    \multicolumn{1}{c|}{} &
    \multicolumn{2}{c|}{PIMC} &
    \multicolumn{2}{c}{SPM} \\
    \hline
    \hline
 2	&	\multicolumn{1}{c|}{--}	& \multicolumn{1}{c|}{--}	&	 -15379	&	143	\\    
 4	&	\multicolumn{1}{c|}{--}	& \multicolumn{1}{c|}{--}	&	-1427	&	29	\\    
 6	&	\multicolumn{1}{c|}{--}	& \multicolumn{1}{c|}{--}	&	 564	&	4	\\    
 8 	&	\multicolumn{1}{c|}{--}	& \multicolumn{1}{c|}{--}	&	1010.0	&	2.0	\\        
10	&	1091.79	&	1.84	&	1092.62	&	1.18	\\
15	&	977.01	&	1.02	&	977.00	&	0.73	\\
20	&	818.25	&	0.53	&	818.26	&	0.46	\\
30	&	589.00	&	0.28	&	588.99	&	0.31	\\
40	&	446.99	&	0.20	&	447.02	&	0.22	\\
50	&	353.06	&	0.14	&	353.10	&	0.16	\\
75	&	218.38	&	0.11	&	218.38	&	0.10	\\
100	&	147.69	&	0.08	&	147.72	&	0.08	\\
150	&	76.24	&	0.07	&	76.26	&	0.06	\\
200	&	41.23	&	0.05	&	41.22	&	0.05	\\
273.16	&	14.31	&	0.04	&	14.33	&	0.04	\\
300	&	8.08	&	0.04	&	8.07	&	0.04	\\
400	&	-6.91	&	0.04	&	-6.89	&	0.04	\\
500	&	-14.89	&	0.03	&	-14.88	&	0.04	\\
750	&	-23.50	&	0.03	&	-23.50	&	0.06	\\
1000	&	-26.29	&	0.02	&	-26.31	&	0.11	\\
    \hline
  \end{tabular}
  \label{tab:RTg_He3}
\end{table}

\section{Conclusions}
Recent improvements in the pair potential~\cite{u2_2020} and especially in the three-body
potential~\cite{u3_2023} for helium have allowed us to reduce the uncertainty of the third virial
coefficient $C(T)$ calculated from first principles by approximately a factor of 4--5.  These
uncertainties are much smaller than those that can be obtained from even the best experiments; for
example, the improvement over experiment is more than two orders of magnitude near 300 K.  It is
clear that first-principles values have completely supplanted experiment for $C(T)$ of both $^4$He
and $^3$He.

An improved propagator and other computational improvements for PIMC have allowed the temperature
range of first-principles $C(T)$ to be extended down to 0.5~K (compared to a previous lower limit of
2.6~K). To our knowledge, these are the first high-accuracy values of $C(T)$, either theoretical or
experimental, to be obtained at such low temperatures.

For the third acoustic virial coefficient $\gamma_\mathrm{a}(T)$, a new path-integral formula was
employed that reduces the statistical uncertainty compared to the previous approach.  However, even
this improved approach produces undesirably large statistical uncertainties below about 50~K.  In
addition, exchange effects have not been implemented in the path-integral calculation of
$\gamma_\mathrm{a}(T)$, limiting its applicability to temperatures above approximately 6~K.
Finally, because $\gamma_\mathrm{a}(T)$ involves temperature derivatives of $C(T)$, it is not clear
how to propagate the uncertainties of the potentials to an uncertainty in the acoustic virial
coefficient.

We therefore implemented for the first time in such calculations the Schlessinger Point Method
(SPM), a statistical approach that uses the $C(T)$ data and their uncertainties to obtain quantities
involving derivatives and, most importantly, a reasonable estimate of the uncertainty of these
quantities.  The SPM is able to produce $\gamma_\mathrm{a}(T)$ and its uncertainty throughout the
range of $C(T)$ calculations.

There are several avenues for further reduction in the uncertainty, depending on the temperature
range, as can be seen in the uncertainty budget shown in Fig.~\ref{fig:uC_He4}.  At low
temperatures (below about 2~K), the largest contribution to our uncertainty budget is the
statistical uncertainty in the PIMC calculations, which could be reduced at the expense of more
computing time.  At somewhat higher temperatures, up to about 10~K, the two-body potential is the
largest source of uncertainty. Above 10 K, the largest uncertainty contribution comes from
uncertainty in the three-body potential.

Finally, we note that in some acoustic applications at higher pressures the fourth acoustic virial
coefficient could be of interest.  Direct PIMC calculation of this quantity would be quite
difficult.  However, the SPM method used here could be applied to the recent first-principles
results for the fourth density virial coefficient $D(T)$ of helium\cite{Wheatley_2023} to provide
the needed information for the fourth acoustic virial coefficient.

\begin{acknowledgments}
G.G. acknowledges support from {\em Real-K} project 18SIB02,
which has received funding from the EMPIR programme co-financed by the
Participating States and from the European Union's Horizon 2020 research and
innovation programme.
G.G. acknowledges CINECA (Award No. IscraC-RAVHE) under the ISCRA
initiative for the availability of high-performance computing resources and
support, and the University of Trento for a generous allocation of computing
time.
\end{acknowledgments}

\appendix
\include{supplementary}

\bibliography{main}

\end{document}

%% file: supplementary.tex
\section{Small variance estimation of acoustic virial coefficients}

In the following derivations it will be convenient to evaluate derivatives
with respect to the inverse temperature $\beta = (\kB T)^{-1}$, so that one
has
\begin{eqnarray}
  T \frac{\diff}{\diff T} &=& -\beta \frac{\diff}{\diff\beta} \\
  T^2 \frac{\diff^2}{\diff T^2} &=& 2 \beta \frac{d}{\diff\beta} + \beta^2 \frac{\diff^2}{\diff\beta^2}.
\end{eqnarray}

\subsection{Second acoustic virial}

The second acoustic virial coefficient, $\ba(T)$, is defined as
\begin{equation}
  \ba(T) = 2 B(T) + 2(\gamma_0-1) T \frac{\diff B}{\diff T} +
  \frac{(\gamma_0-1)^2}{\gamma_0} T^2 \frac{\diff^2B}{\diff T^2},
  \label{eq:beta_a}
\end{equation}
where $B(T)$ is the density virial coefficient and in the case of a
monoatomic system, $\gamma_0 = 5/3$.

In the path-integral approximation, the expression of the second virial
coefficient $B(T)$ can be written
as
\begin{eqnarray}
B(T) &=& -\frac{N_\mathrm{A}}{2} \int \left(
\e^{-\beta v(X)}-1
\right) F(X) ~ \diff X \\
&\equiv& -\frac{N_\mathrm{A}}{2} \int b(X) F(X) ~ \diff X, \label{eq:b}
\end{eqnarray}
where $X$ denotes the $6P-3$ cartesian coordinates $x_i$ ($i=1,\ldots,6P-3$) needed to describe a
pair of ring polymers, $F(X)$ is the distribution function of ring polymer
configurations and $v(X)$ is the pair potential averaged over the ring
polymers.~\cite{Garberoglio2009b,Garberoglio2011,Garberoglio2011a}
Equation~(\ref{eq:b}) is the definition of the quantity $b(X)$.
The quantity $F(X)$ can be written as $A \e^{-\beta U}$, where $U$ has the
form of a harmonic potential and $A$ is a (temperature dependent)
proportionality constant. Given the expression of
$F(X)$,~\cite{Garberoglio2009b,Garberoglio2011,Garberoglio2011a} it is
straighforward to verify that
\begin{equation}
  \frac{\diff F}{\diff\beta} = \left( U - \frac{3(P-1)}{2\beta} \right) F,
\label{eq:dF}
\end{equation}
and hence
\begin{equation}
  \frac{\diff B}{\diff\beta} = -\frac{N_\mathrm{A}}{2} \int
  - v \e^{-\beta v} F +
  \left(\e^{-\beta v} - 1\right) \left(U - \frac{3(P-1)}{\beta}\right) F  
  ~ \diff X.
\label{eq:dBdbeta}    
\end{equation}

Notice that it is the presence of the terms coming from Eq.~(\ref{eq:dF}) in
Eq.~(\ref{eq:dBdbeta}), which are similar to the {\em thermodynamic}
estimator of the kinetic energy in the path-integral
approach,~\cite{Tuckerman10} that are responsible for the large statistical
variance in our earlier calculation of acoustic
virials.~\cite{Garberoglio2011}

The fact that $U(X)$ has the form of a harmonic potential -- {\em i.e.}, is
a homogeneous function of degree two -- enables the application of Euler's
theorem which results in
\begin{equation}
  2U = \sum_{i=1}^{6P-3} x_i \frac{\partial U}{\partial x_i}.
  \label{eq:Euler}
\end{equation}
This result is at the heart of the derivation of the {\em virial} estimator
of the kinetic energy.~\cite{Herman82}
In fact, one can write
\begin{equation}
\frac{\partial U}{\partial x_i} \e^{-\beta U} = -\frac{1}{\beta} \frac{\partial}{\partial
  x_i}\e^{- \beta U},
\end{equation}
and integrate by parts Eq.~(\ref{eq:dBdbeta}) resulting in the expression
\begin{equation}
\frac{\diff B}{\diff\beta} = -\frac{N_\mathrm{A}}{2} \int F
\underbrace{\left[ \frac{3}{2 \beta} (\e^{-\beta v} - 1) - W \e^{-\beta v}
    \right]}_{\equiv Q} ~ \diff X,
\label{eq:dBvirial}
\end{equation}
where
\begin{equation}
W = v + \sum_{i} \frac{x_i}{2}\frac{\partial v}{\partial x_i}.  
\end{equation}
At this point, we can evaluate the second derivative with respect to
$\beta$ of $B(T)$, obtaining
\begin{equation}
  \frac{\diff^2B}{\diff\beta^2} =
  -\frac{N_\mathrm{A}}{2} \int \frac{\diff F}{\diff\beta} Q + F \left( W v \e^{-\beta v} -
  \frac{3}{2\beta^2} (\e^{-\beta v} - 1) - \frac{3 v}{2 \beta} \e^{-\beta
    v}\right) ~ \diff X.  
\end{equation}
In this case, the largest contribution to the variance comes from the first
term, which we can again integrate by parts obtaining
\begin{equation}
\int \frac{\diff F}{\diff\beta} Q = \frac{1}{2\beta} \int F \left( 3Q + 
\sum_i x_i \frac{\partial Q}{\partial x_i} \right) ~ \diff X,  
\end{equation}
with
\begin{eqnarray}
\frac{\partial W}{\partial x_i} &=& \frac{3}{2} \frac{\partial v}{\partial
  x_i} + \frac{1}{2} \sum_{k} x_k \frac{\partial^2 v}{\partial x_i \partial
  x_k} \\
\frac{\partial Q}{\partial x_i} &=& \e^{-\beta v} \left[ -\frac{\partial
  W}{\partial x_i} + \frac{\partial v}{\partial x_i} \left( \beta W -
  \frac{3}{2}\right) \right],
\end{eqnarray}
so that putting all together
\begin{widetext}
\begin{eqnarray}
  \frac{\diff^2 B}{\diff\beta^2} &=& -\frac{N_\mathrm{A}}{2}
  \int \left[
  \frac{3 \beta Q}{2} + \frac{1}{2} \sum_i x_i \frac{\partial \beta
    Q}{\partial x_i}
  + \left(\beta v \beta W - \frac{3 \beta v}{2}\right) \e^{-\beta v} - \frac{3}{2}
  (\e^{-\beta v} - 1)
  \right] F ~ \diff X \\
  &=&
  -\frac{N_\mathrm{A}}{2} \int \left[
    \frac{3}{4} \left(\e^{-\beta v}-1\right) +
    \beta v \beta W \e^{-\beta v} + \right. \nonumber \\
    & & \left.
    \frac{\beta \e^{-\beta v}}{2} \left(
      (\beta W - \frac{3}{2}) \mbX \cdot \nabla v  - 6 W
    -\frac{1}{2} \mbX \cdot \nabla^2 v \cdot \mbX \right)
    \right] F ~ \diff X \\
  &\equiv&
  -\frac{N_\mathrm{A}}{2} \int S(X) ~ \diff X,
  \label{eq:d2Bvirial}
\end{eqnarray}  
\end{widetext}
where the last equation is the definition of $S(X)$.

Finally, we notice that due to translational and rotational
invariance we can write
\begin{eqnarray}
  \mbX \cdot \nabla v &=& \sum_{i=1}^{6P-3} x_i \frac{\partial v}{\partial
    x_i} = \sum_{k=1}^P r^{(k)} \frac{\partial v}{\partial r^{(k)}}
  \label{eq:rot1}\\
  \mbX \cdot \nabla^2 v \cdot \mbX &=& \sum_{ij} x_i x_j \frac{\partial^2
    v}{\partial x_i x_j} = \sum_{k=0}^P {r^{(k)}}^2 \frac{\partial^2
    v}{\partial {r^{(k)}}^2},
  \label{eq:rot2}
\end{eqnarray}
where $r^{(k)}$ is the distance between the two atoms in the imaginary-time
slice $k$.

Equations (\ref{eq:beta_a}), (\ref{eq:dBvirial}), and (\ref{eq:d2Bvirial})
enable the evaluation of the second acoustic virial coefficient with
a variance greatly reduced with respect to the straightforward application
of temperature derivatives.

\subsection{Third acoustic virial}

The expression for the third acoustic virial coefficient is a bit more complicated~\cite{Gillis96}
\begin{eqnarray}
  Q &=& B + (2\gamma_0-1) T \frac{\diff B}{\diff T} + (\gamma_0-1) T^2 \frac{\diff^2B}{\diff T^2} \\
  \RTg &=& \left(\frac{\gamma_0 -1}{\gamma_0} Q^2 - \ba(T) B(T)\right) + 
  \frac{2 \gamma_0 + 1}{\gamma_0} C + \nonumber \\
  & & \frac{\gamma_0^2-1}{\gamma_0} T\frac{\diff C}{\diff T} + \frac{(\gamma_0 -1)^2}{2
    \gamma_0} T^2 \frac{\diff^2 C}{\diff T^2}, \label{eq:RTg_a}
\end{eqnarray}
but we can easily recognize a series of terms involving only two bodies
(coming from $Q$, $\ba$, and the two-body part of $C(T)$) that can be
evaluated using the approach outlined in the previous section.

\subsubsection{The two-body contribution}

We define the two-body contribution to $C(T)$ as
\begin{equation}
  C_{2\mathrm{B}}(T) = \int F(X) F(Y) \left(b(X) - 1 \right) \left(b(Y) - 1 \right) ~ \diff X \diff Y,
\end{equation}
so that, from Eq.~(\ref{eq:RTg_a}), the two-body contribution to the third acoustic virial is then
\begin{widetext}
\begin{equation}
  \RTg^{(2\mathrm{B})} = \left(\frac{\gamma_0 -1}{\gamma_0} Q^2 - \ba(T) B(T)\right) + 
  \frac{2 \gamma_0 + 1}{\gamma_0} C_{2\mathrm{B}} +
  \frac{\gamma_0^2-1}{\gamma_0} T\frac{\diff C_{2\mathrm{B}}}{\diff T} + \frac{(\gamma_0 -1)^2}{2
    \gamma_0} T^2 \frac{\diff^2 C_{2\mathrm{B}}}{\diff T^2}, \label{eq:RTg_2B}
\end{equation}
  
\end{widetext}
which is most conveniently evaluated by defining
\begin{eqnarray}
  b_T(X) &=& -\beta Q(X) \\
  b_{TT}(X) &=& \beta^2 S(X) - 2 b_T(X).
\end{eqnarray}

After some lenghty but straightforward calculations, one ends up with
\begin{widetext}
\begin{eqnarray}
  \RTg^{(2\mathrm{B})} &=& \int \left(
    \frac{2}{15}b_{TT}(X)b_{TT}(Y) +
    \frac{14}{15} b_T(X) b_{TT}(Y) +
     b(X) b_{TT}(Y) + \right. \nonumber \\
    & & \left. \frac{73}{30} b_T(X) b_T(Y) +    
    \frac{34}{5} b(X) b_T(Y) +
    \frac{33}{5} b(X) b(Y) \right) F(X) F(Y) ~ \diff X \diff Y.
\end{eqnarray}  
\end{widetext}

\subsubsection{The three-body contribution}

Additionally, $\RTg(T)$ has a contribution from a purely three-body term, coming from the corresponding
contribution to $C(T)$
\begin{equation}
  C_{3\mathrm{B}}(T) = -\frac{1}{3} \int F \left(Z_3 - 3Z_2 + 2\right) ~ \diff X,
\end{equation}
that we are going to discuss in some detail, following the same steps of
the previous section. The derivative with respect to $\beta$ can be
evaluated with the same integration-by-parts techniques outlined above, obtaining
\begin{equation}
  \frac{\partial C_{3\mathrm{B}}}{\partial \beta} =
  -\frac{1}{\beta} \int F \left(Z_3 - 3Z_2 + 2\right) ~ \diff X
  +  \\
  \frac{1}{3} \int F \left( W_3 Z_3 - 3 W_2 Z_2 \right)~ \diff X,
\end{equation}
where we have used
\begin{eqnarray}
  \frac{\partial F}{\partial \beta} &=& F \left(U - \frac{9 (P-1)}{2
    \beta}\right) \\
  W_3 &=& V_3 + \frac{1}{2} \sum_{i=1}^{9(P-1)} x_i \frac{\partial V_3}{\partial x_i} \\
  W_2 &=& V_2 + \frac{1}{2} \sum_{i}^{6(P-1)} x_i \frac{\partial V_2}{\partial x_i} \\
  U F &=& -\frac{1}{2\beta} \sum_i x_i \frac{\partial F}{\partial x_i},
\end{eqnarray}
where we have denoted by $V_3$ the total (additive + non-additive)
three-body potential and by $V_2$ a two-body interaction.
The second derivative with respect to $\beta$ becomes:
\begin{eqnarray}
  \frac{\partial^2 C_{3\mathrm{B}}}{\partial \beta^2} &=&
  -\frac{2}{\beta^2} \int F \left( Z_3 - 3 Z_2 + 2 \right)~ \diff X + \nonumber \\  
  & & \frac{1}{3 \beta} \int F \left[ Z_3 W_3 \left(\frac{6}{\beta} - W_3\right) -
    3 Z_2 W_2 \left(\frac{6}{\beta} - W_2\right) \right] ~ \diff X + \nonumber \\
  & &   \frac{1}{6\beta} \int F \left[
    Z_3 \left( \frac{1}{2} \mbX \cdot \nabla^2 V_3 \cdot \mbX +
    \frac{3}{2} \mbX \cdot \nabla V_3 \right) - \right. \nonumber \\
    & & \left.
    3 Z_2 \left( \frac{1}{2} \mbX \cdot \nabla^2 V_2 \cdot \mbX +
    \frac{3}{2} \mbX \cdot \nabla V_2 \right) \right] ~ \diff X.
\end{eqnarray}  

Using again translational and rotational invariance, and defining $R_1 =
r_{12}$, $R_2=r_{13}$ and $R_3 = r_{23}$, we can write
\begin{equation}
  \mbX \cdot \nabla V_3 = \sum_{k=0}^P \sum_{i=0}^3
  R^{(k)}_i \frac{\partial V_3}{\partial R^{(k)}_i},
\end{equation}
and
\begin{equation}
  \mbX \cdot \nabla^2 V_3 \cdot \mbX = \sum_{k=0}^P  
  \sum_{ij=0}^3 R^{(k)}_i R^{(k)}_j \frac{\partial^2 V_3}{\partial
    R^{(k)}_i \partial R^{(k)}_j},
\end{equation}
where the superscript $(k)$ denotes an imaginary-time slice.
Terms involving the derivatives of $V_2$ can be evaluated using
Eqs.~(\ref{eq:rot1}) and (\ref{eq:rot2}).

\section{Statistical Schlessinger Point Method}

\subsection{The second acoustic virial coefficient for ${}^4$He}

\begin{figure}[h]
  \center\includegraphics[width=0.8\linewidth]{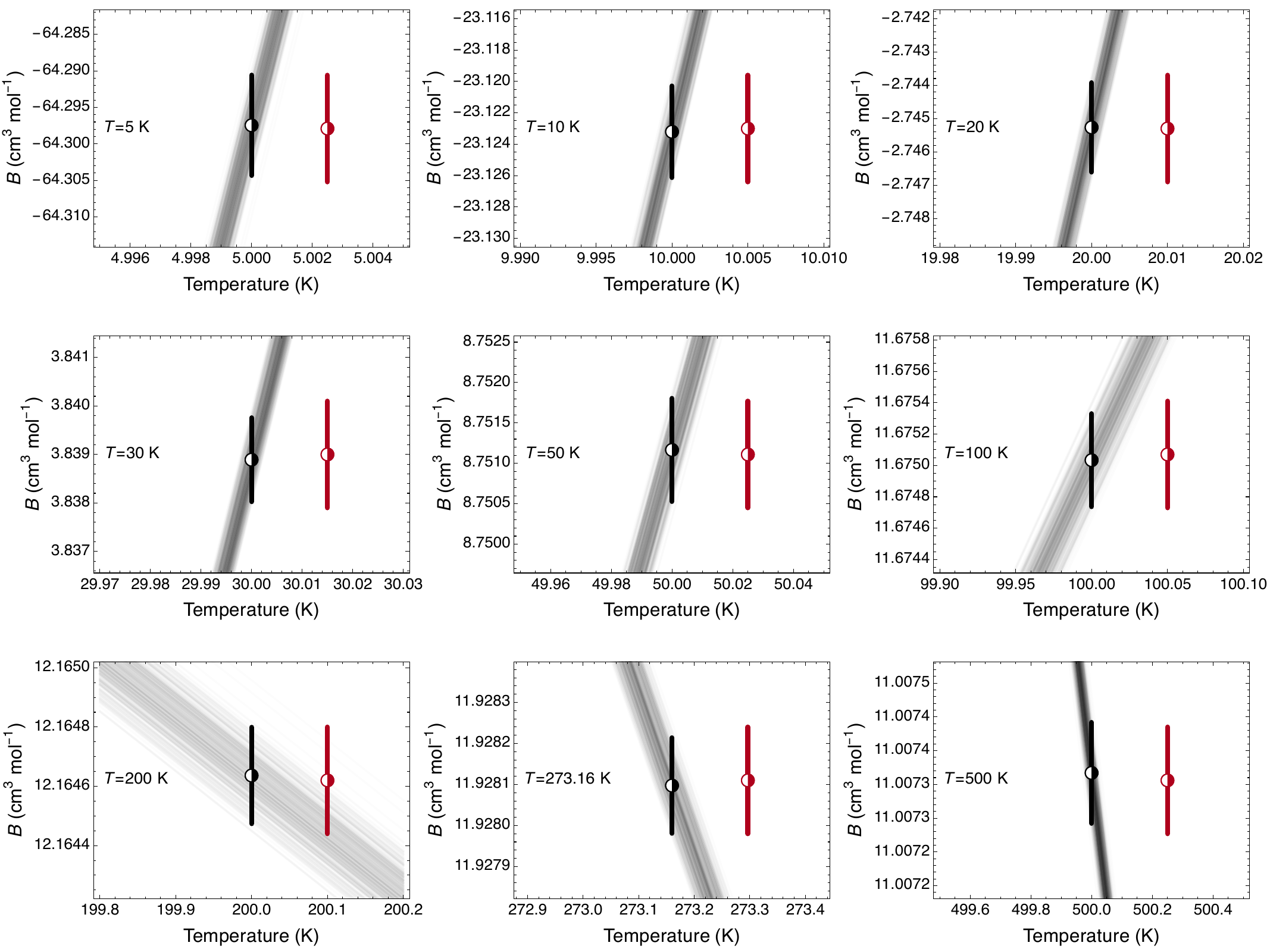}  
  \caption{Interpolating curves generated using the SPM method in the case of $B(T)$ for
    ${}^4$He at various temperatures. The black symbol in each panel denotes the
    average value and $k=2$ expanded uncertainty derived from the SPM curves. The red symbol, which
    has been displaced to the right for the sake of clarity, reports the average value and expanded
    uncertainty from the original calculation.~\cite{u2_2020} }
\label{fig:B_SPM_a}  
\end{figure}

Figure~\ref{fig:B_SPM_a} reports the values of $B(T)$ reconstructed using the SPM approach, and
compares them with the direct calculation obtained using the phase-shift approach.~\cite{u2_2020}

\begin{figure}[h]
  \center\includegraphics[width=0.8\linewidth]{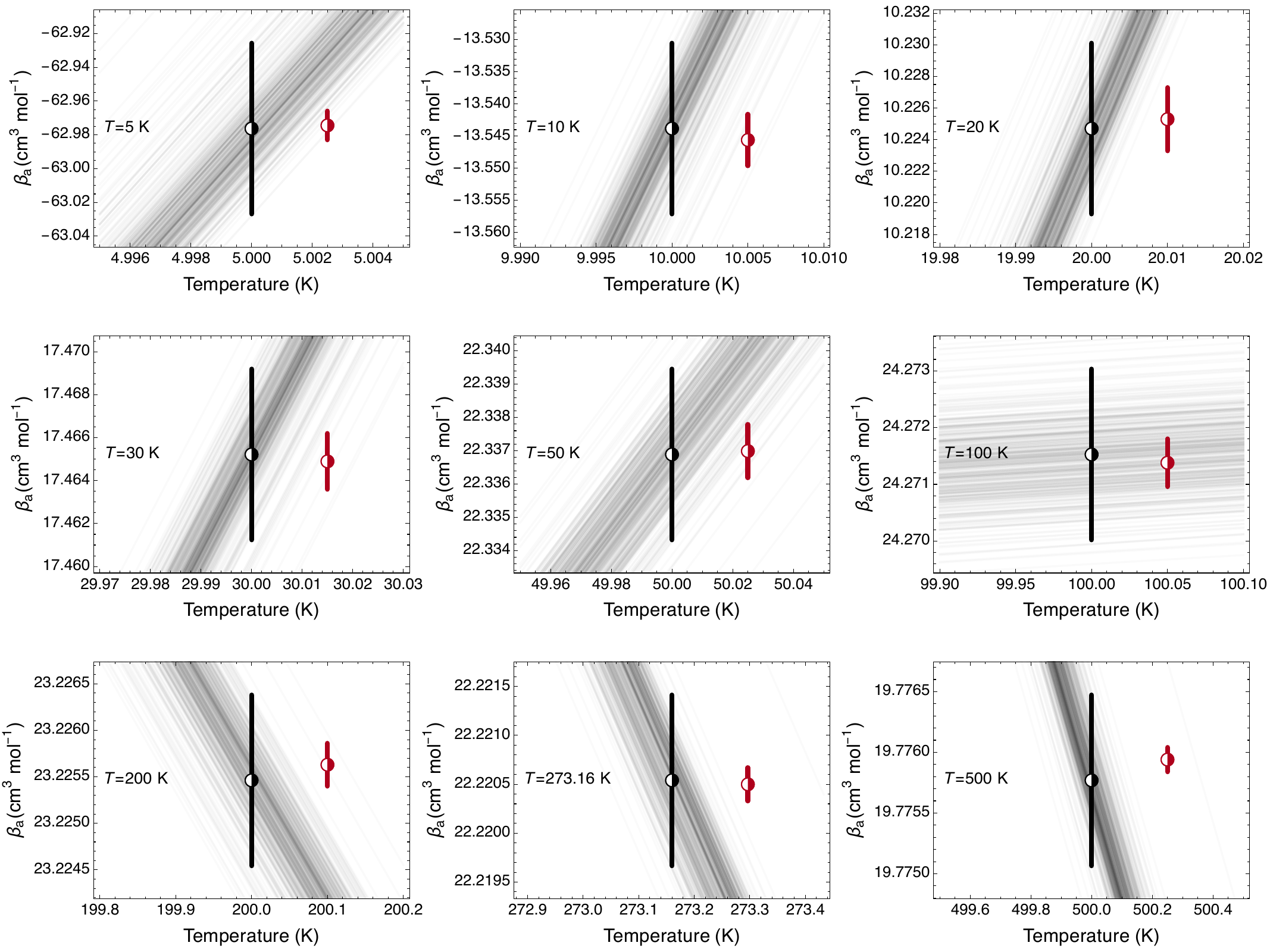}  
  \caption{Interpolating curves generated using the SPM method in the case of $\ba(T)$ for
    ${}^4$He at various temperatures. The black symbol in each panel denotes the
    average value and $k=2$ expanded uncertainty derived from the SPM curves. The red symbol, which
    has been displaced to the right for the sake of clarity, reports the average value and expanded
    uncertainty from the original calculation.~\cite{u2_2020} }
\label{fig:beta_SPM_a}  
\end{figure}

Figure~\ref{fig:beta_SPM_a} reports the values of $\ba(T)$ reconstructed using the SPM approach, and
compares them with the direct calculation obtained using the phase-shift approach.~\cite{u2_2020}

\subsection{The third acoustic virial coefficient for ${}^4$He}

\begin{figure}
  \center\includegraphics[width=0.8\linewidth]{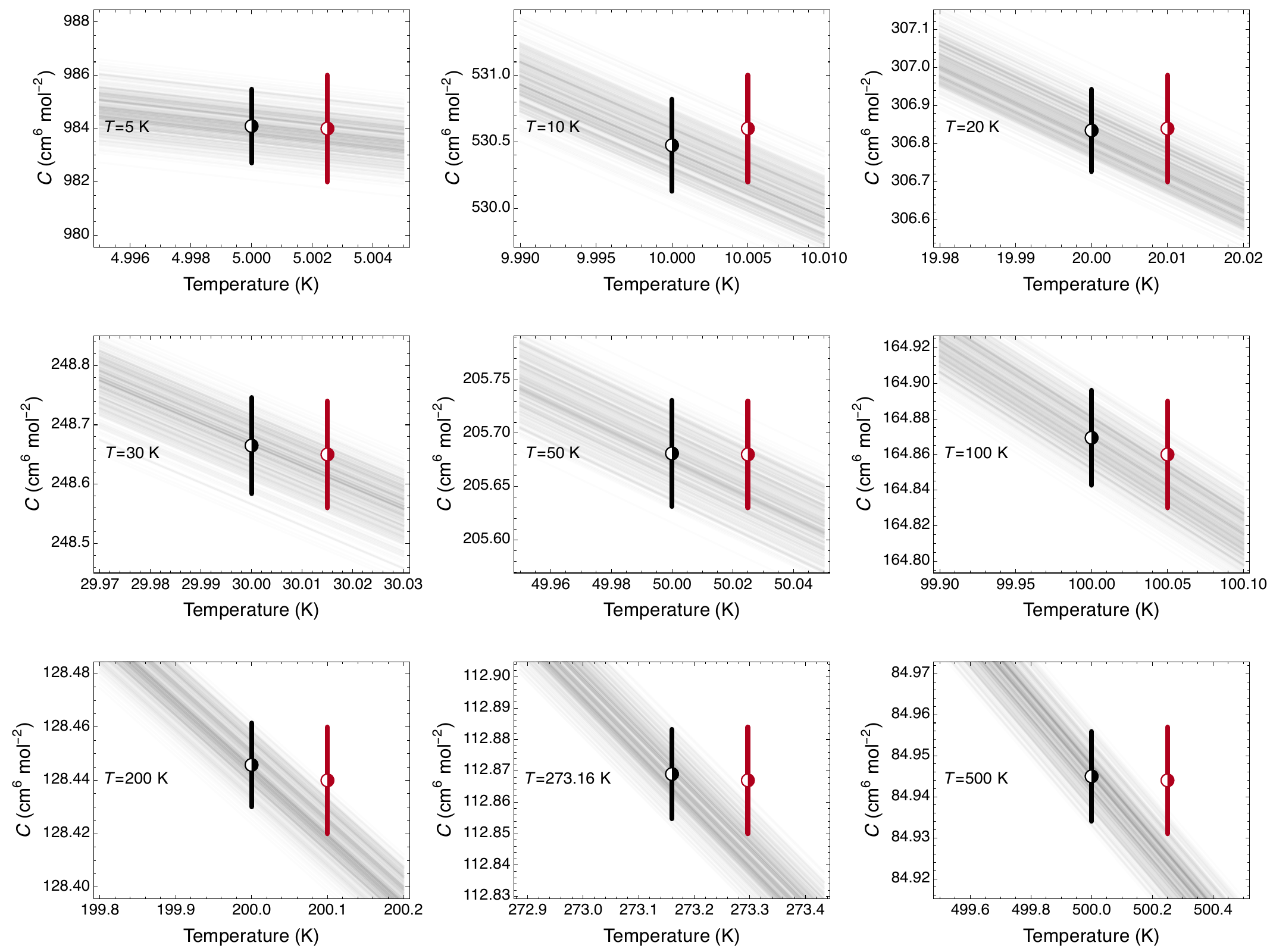}  
  \caption{Interpolating curves generated using the SPM method in the case of $C(T)$ for
    ${}^4$He at various temperatures. The black symbol in each panel denotes the
    average value and $k=2$ expanded uncertainty derived from the SPM curves. The red symbol, which
    has been displaced to the right for the sake of clarity, reports the average value and expanded
    uncertainty from path-integral Monte Carlo calculations performed in this work.}
\label{fig:C_SPM}  
\end{figure}

Figure~\ref{fig:C_SPM} reports the values of $C(T)$ reconstructed using the
SPM approach, and compares them with the direct calculation obtained using
the path-integral approach in this work.

\begin{figure}
  \center\includegraphics[width=0.8\linewidth]{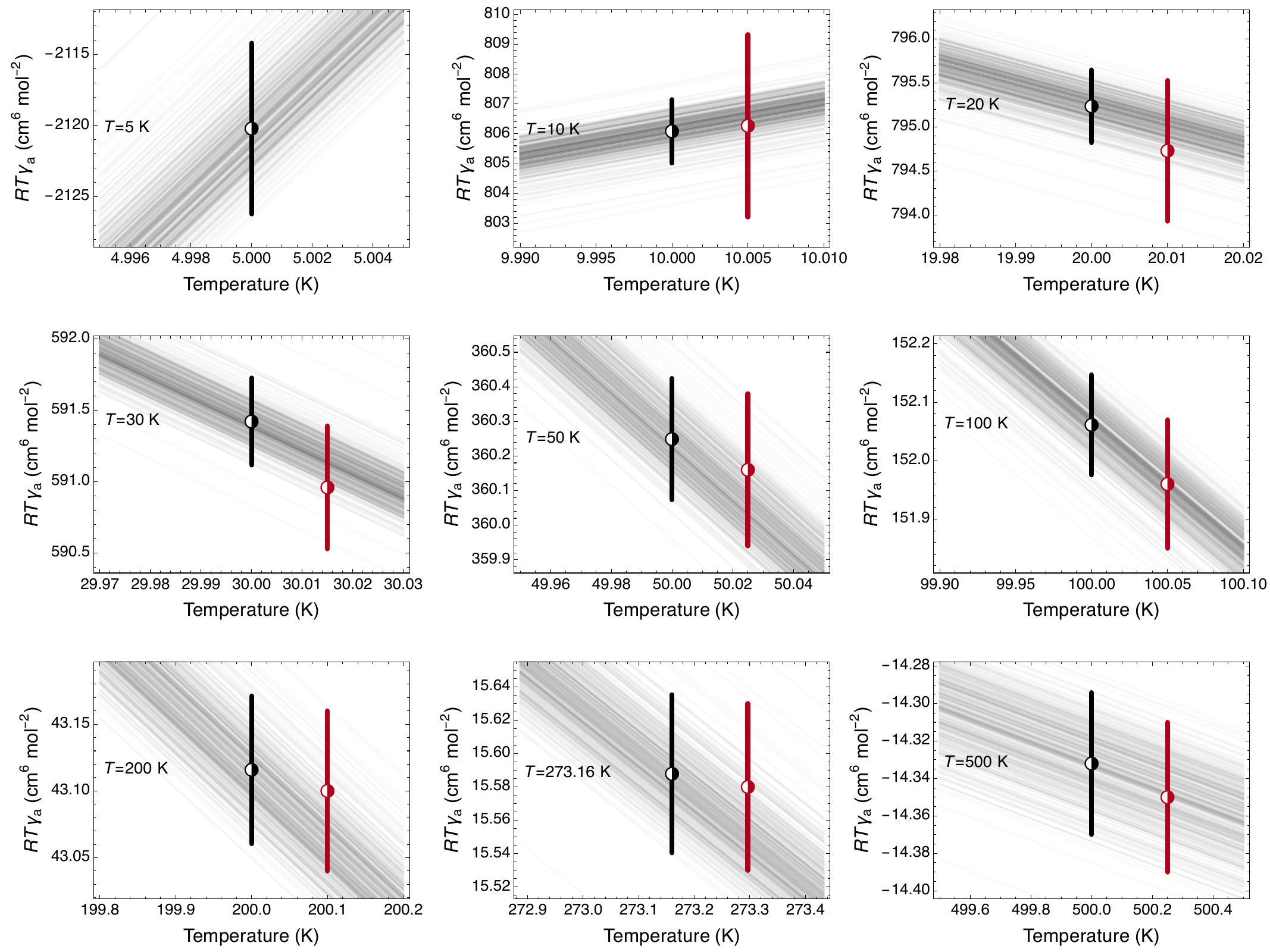}  
  \caption{Interpolating curves generated using the SPM method in the case of $\RTg(T)$ for
    ${}^4$He at various temperatures. The black symbol in each panel denotes the
    average value and $k=2$ expanded uncertainty derived from the SPM curves. The red symbol, which
    has been displaced to the right for the sake of clarity, reports the average value and expanded
    uncertainty from path-integral Monte Carlo calculations, which were performed in
    this work only for $T \geq 10$~K.}
\label{fig:RTg_SPM}  
\end{figure}

Figure~\ref{fig:RTg_SPM} reports the values of $\RTg (T)$ reconstructed using the SPM approach, and
compares them with the direct calculation obtained using  path-integral approach in this work.

